\documentclass[sigplan,authorversion,nonacm,screen,twocolumn]{acmart}
\acmSubmissionID{766}

\setcopyright{none}

\hypersetup{breaklinks=true}



\usepackage[utf8]{inputenc}
\usepackage[T1]{fontenc}

\usepackage{newtxtext,newtxmath}
\usepackage{booktabs}
\usepackage{threeparttable}

\usepackage[english]{babel}

\usepackage{glossaries}
\glsdisablehyper

\def \txone {ThunderX\protect\nobreakdash-1\xspace}

\def \txoneabbr {\textsc{tx1}\xspace}

\def \tz {Arm TrustZone\xspace}

\def \ryzen {AMD Ryzen\xspace}

\def \ryzenabbr {\textsc{ryzen}\xspace}

\def \aspeed {AST~2600\xspace}
\def \aspeedlong {ASPEED AST~2600\xspace}
\def \aspeedabbr {\textsc{aspeed}\xspace}

\def \jetson {Jetson\xspace}
\def \jetsonlong {Nvidia Jetson~TX2\xspace}
\def \jetsonabbr {\textsc{jetson}\xspace}

\def \pico {Pico~2\xspace}
\def \picolong {Raspberry Pi Pico 2\xspace}
\def \picoabbr {\textsc{pico}\xspace}

\def \stm {STM32H753\xspace}
\def \stmlong {STMicro 32H753\xspace}
\def \stmabbr {\textsc{stm}\xspace}

\def \omap {OMAP\xspace}
\def \omaplong {Texas Instruments OMAP 4460\xspace}
\def \omapabbr {\textsc{omap}\xspace}

\def \zynq {XU5\xspace}
\def \zynqlong {Xilinx Zynq UltraScale+ XU5\xspace}
\def \zynqabbr {\textsc{zynq}\xspace}

\newacronym{acpi}{ACPI}{Advanced Configuration and Power Interface}
\newacronym{acs}{ACS}{ATA Command Set}
\newacronym{afu}{AFU}{Accelerator Function Unit}
\newacronym{ahci}{AHCI}{Advanced Host Controller Interface}
\newacronym{asc}{ASC}{Address Space Controller}
\newacronym{ascii}{ASCII}{American Standard Code for Information Interchange}
\newacronym{asic}{ASIC}{Application-Specific Integrated Circuit}
\newacronym{ata}{ATA}{AT Attachment}
\newacronym{atapi}{ATAPI}{AT Attachment Packet Interface}
\newacronym{atf}{ATF}{ARM Trusted Firmware}
\newacronym{bar}{BAR}{Base Address Register}
\newacronym{bdk}{BDK}{Board Development Kit}
\newacronym{bist}{BIST}{Built-in Self-test}
\newacronym{bmc}{BMC}{Baseboard Management Controller}
\newacronym{bram}{BRAM}{Block RAM}
\newacronym{caam}{CAAM}{Cryptographic Acceleration and Assurance Module}
\newacronym{cam}{CAM}{Common Access Method}
\newacronym{capi}{CAPI}{Coherent Accelerator Processor Interface}
\newacronym{capp}{CAPP}{Coherent Accelerator Processor Proxy}
\newacronym{cca}{CCA}{Confidential Compute Architecture}
\newacronym{ccip}{CCI-P}{Core Cache Interface}
\newacronym{ccix}{CCIX}{Cache Coherent Interconnect for Accelerators}
\newacronym{ccpi}{CCPI}{Cavium Coherent Processor Interconnect}
\newacronym{ci}{CI}{Continuous Intergration}
\newacronym{clut}{CLUT}{Cache Line Under Test}
\newacronym{cpld}{CPLD}{Complex Programmable Logic Device}
\newacronym{cpu}{CPU}{Central Processing Unit}
\newacronym{cxl}{CXL}{Compute eXpress Link}
\newacronym{dac}{DAC}{Digital Analog Converter}
\newacronym{dc}{DC}{Display Controller}
\newacronym{dep}{DEP}{Data Execution Prevention}
\newacronym{dma}{DMA}{Direct Memory Access}
\newacronym{dos}{DoS}{denial-of-service}
\newacronym{dsl}{DSL}{domain-specific language}
\newacronym{dsp}{DSP}{Digital Signal Processor}
\newacronym{eci}{ECI}{Enzian Coherence Interface}
\newacronym{edma}{eDMA}{Enhanced Direct Memory Access}
\newacronym{ept}{EPT}{Extended Page Table}
\newacronym{esai}{ESAI}{Enhanced Synchronous Audio Interface}
\newacronym{esp}{ESP}{Executable Space Protection}
\newacronym{fat}{FAT}{File Allocation Table}
\newacronym{fis}{FIS}{Frame Information Structure}
\newacronym{fiu}{FUI}{FPGA Interface Unit}
\newacronym{fmc}{FMC}{FPGA Mezzanine Card}
\newacronym{fpga}{FPGA}{Field-Programmable Gate Array}
\newacronym{fsis}{FSIS}{Filesystem Information Sector}
\newacronym{fsm}{FSM}{Finite State Machine}
\newacronym{gpu}{GPU}{Graphics Processing Unit}
\newacronym{gqr}{GQR}{Generalized-Quasi-Reduction}
\newacronym{gsync}{\textsc{GSync}}{Global Synchronization}
\newacronym{hba}{HBA}{Host Bus Adapter}
\newacronym{hbm}{HBM}{High-Bandwidth Memory}
\newacronym{hpc}{HPC}{High-Performance Computing}
\newacronym{i2c}{I\textsuperscript{2}C}{Inter-Integrated Circuit}
\newacronym{ich}{ICH}{I/O Controller Hub}
\newacronym{ic}{IC}{Integrated Circuit}
\newacronym{idc}{IDC}{Inter-Domain Communication}
\newacronym{iee}{IEE}{Inline Encryption Engine}
\newacronym{iommu}{IOMMU}{Input-Output Memory Management Unit}
\newacronym{iot}{IoT}{Internet-of-Things}
\newacronym{ipi}{IPI}{Inter-Processor Interrupt}
\newacronym{ip}{IP}{intellectual property block}
\newacronym{ipmi}{IPMI}{Intelligent Platform Management Interface}
\newacronym{isa}{ISA}{instruction-set architecture}
\newacronym{iut}{IUT}{Implementation Under Test}
\newacronym{kaslr}{KASLR}{Kernel Address Space Layout Randomization}
\newacronym{l2c}{L2C}{Second-Level Cache}
\newacronym{lba}{LBA}{Logical Block Address}
\newacronym{llc}{LLC}{Last-Level Cache}
\newacronym{lru}{LRU}{Least-recently used}
\newacronym{mmu}{MMU}{Memory Management Unit}
\newacronym{mpsoc}{MPSoC}{Multiprocessor System-on-a-Chip}
\newacronym{mpu}{MPU}{Memory Protection Unit}
\newacronym{mpx}{MPX}{Memory Protection Extensions}
\newacronym{msi}{MSI}{Message-Signalled Interrupt}
\newacronym{mu}{MU}{Messaging Unit}
\newacronym{ncq}{NCQ}{Native Command Queueing}
\newacronym{nic}{NIC}{Network Interface Adaptor}
\newacronym{numa}{NUMA}{Non-Uniform Memory Access}
\newacronym{nvme}{NVMe}{NVM Express}
\newacronym{ocapi}{OpenCAPI}{Open Coherent Accelerator Processor Interface}
\newacronym{oci}{OCI}{Octeon III multi-node Interconnect}
\newacronym{os}{OS}{Operating System}
\newacronym{pae}{PAE}{Physical Address Extensions}
\newacronym{pata}{PATA}{Parallel ATA}
\newacronym{pcb}{PCB}{Printed Circuit Board}
\newacronym{pcie}{PCIe}{PCI Express}
\newacronym{pci}{PCI}{Peripheral Component Interconnect}
\newacronym{piix}{PIIX}{PCI IDE ISA Xcelerator}
\newacronym{pio}{PIO}{Programmed I/O}
\newacronym{pmbus}{PMBus}{Power Management Bus}
\newacronym{prd}{PRD}{Physical Region Descriptor}
\newacronym{psci}{PSCI}{Power State Coordination Interface}
\newacronym{psl}{PSL}{POWER Service Layer}
\newacronym[longplural={Page Table Entries}]{pte}{PTE}{Page Table Entry}
\newacronym{qpi}{QPI}{QuickPath Interconnect}
\newacronym{rdma}{RDMA}{Remote Direct Memory Access}
\newacronym{rfis}{RFIS}{Received FIS}
\newacronym{rpc}{RPC}{Remote Procedure Call}
\newacronym{rtc}{RTC}{Real Time Clock}
\newacronym{rtl}{RTL}{Register-transfer level}
\newacronym{sai}{SAI}{Synchronous Audio Interface}
\newacronym{sata}{SATA}{Serial ATA}
\newacronym{sbc}{SBC}{single board computer}
\newacronym{scsi}{SCSI}{Small Computer System Interface}
\newacronym{scu}{SCU}{System Controller Unit}
\newacronym{seco}{SECO}{Security Controller}
\newacronym{sgx}{SGX}{Software Guard Extensions}
\newacronym{simd}{SIMD}{Single Input Multiple Data}
\newacronym{sim}{SIM}{SCSI Interface Module}
\newacronym{skb}{SKB}{System Knowledgebase}
\newacronym{smbus}{SMBus}{System Management Bus}
\newacronym{smc}{SMC}{Secure Monitor Call}
\newacronym{smi}{SMI}{System Management Interrupt}
\newacronym{smm}{SMM}{System Management Mode}
\newacronym{smmu}{SMMU}{System Memory Management Unit}
\newacronym{smt}{SMT}{satisfiability modulo theories}
\newacronym[longplural={Systems-on-Chip}]{soc}{SoC}{System-on-Chip}
\newacronym[longplural={Systems-on-Module}]{som}{SoM}{System-on-Module}
\newacronym{spl}{SPL}{System Protocol Layer}
\newacronym{tap}{TAP}{Test Access Port}
\newacronym{tcb}{TCB}{Trusted Computing Base}
\newacronym{tcm}{TCM}{Tightly Coupled Memory}
\newacronym{tcp}{TCP}{Transmission Control Protocol}
\newacronym{tdp}{TDP}{Thermal Design Power}
\newacronym{tee}{TEE}{Trusted Execution Environment}
\newacronym{tfa}{TF-A}{Trusted Firmware-A}
\newacronym{tlb}{TLB}{Translation Lookaside Buffer}
\newacronym{tpu}{TPU}{Tensor Processing Unit}
\newacronym{ttbr}{TTBR}{Translation Table Base Register}
\newacronym{uart}{UART}{universal asynchronous receiver-transmitter}
\newacronym{uefi}{UEFI}{Unified Extensible Firmware Interface}
\newacronym{upi}{UPI}{Universal Path Interconnect}
\newacronym{usb}{USB}{Universal Serial BUS}
\newacronym{vfpga}{vFPGA}{Virtual FPGA}
\newacronym{vfs}{VFS}{Virtual Filesystem}
\newacronym{vliw}{VLIW}{Very Long Instruction Word}
\newacronym{vpu}{VPU}{Video Processing Unit}
\newacronym{xmpu}{XMPU}{Xilinx Memory Protection Unit}
\newacronym{xppu}{XPPU}{Xilinx Peripheral Protection Unit}
\newacronym{xrdc}{XRDC}{Extended Resource Domain Controller}

\glsunset{os}
\glsunset{fpga}
\glsunset{pci}
\glsunset{pcie}
\glsunset{usb}

\ifdefined\isdiff
\newcommand{\revisionComment}[1]{\textbf{[#1]}}
\newcommand{\shepherdComment}[1]{\textbf{[#1]}}
\usepackage{color}
\usepackage[normalem]{ulem}

\else
\newcommand{\revisionComment}[1]{}
\newcommand{\shepherdComment}[1]{}
\fi

\ifdefined\isdiff
  \includecomment{revisionReqs}
\else
  \excludecomment{revisionReqs}
\fi

\usepackage{listings}
\usepackage{lstlangsockeye}
\NewCommandCopy{\originallstinline}{\lstinline}
\RenewDocumentCommand{\lstinline}{O{}}{%
  \originallstinline[basicstyle=\ttfamily,#1]%
}

\usepackage[nameinlink,noabbrev]{cleveref}

\crefname{section}{\S}{\S\S} %
\Crefname{section}{\S}{\S\S} %
\crefformat{line}{L#2#1#3}
\crefformat{section}{\S#2#1#3}

\usepackage{apptools}
\crefname{appendix}{\IfAppendix{section}{appendix}}{\IfAppendix{sections}{appendices}}
\makeatletter
\AddToHook{cmd/appendix/before}{\def\cref@section@alias{appendix}}
\makeatother

\usepackage[nounderscore]{syntax}

\usepackage{tikz}
\usetikzlibrary{tikzmark}

\usepackage{xspace}
\newcommand{\system}{Sockeye\xspace}

\newcommand{\etal}{\textit{et al.}\xspace}

\newcommand{\sockeyeCommit}[0]{\texttt{faa94a3a}}

\begin{document}

\settopmatter{printfolios=true}
\pagestyle{empty}

\title{Sockeye: Bug-finding \textit{and} proofs for platform configurations and hardware based on reference manuals}

\settopmatter{authorsperrow=4}
\author{Ben Fiedler}
\orcid{0000-0002-7215-9147}
\affiliation{
  \institution{ETH Zürich} \city{Zurich} \country{Switzerland}
}

\author{Sedan Abdelgawad}
\orcid{0009-0000-3567-3211}
\affiliation{
  \institution{ETH Zürich} \city{Zurich} \country{Switzerland}
}

\author{Teymour Aldridge}
\orcid{0009-0001-4807-0476}
\affiliation{
  \institution{ETH Zürich} \city{Zurich} \country{Switzerland}
}

\author{Viktor Fukala}
\orcid{0009-0000-2074-7329}
\affiliation{
  \institution{ETH Zürich} \city{Zurich} \country{Switzerland}
}

\author{Jan Häussermann}
\orcid{0009-0005-2517-0908}
\affiliation{
  \institution{ETH Zürich} \city{Zurich} \country{Switzerland}
}

\author{Gamal Hassan}
\orcid{0009-0000-2083-1459}
\affiliation{
  \institution{ETH Zürich} \city{Zurich} \country{Switzerland}
}

\author{Lars Leuthold}
\orcid{0009-0006-1735-2717}
\affiliation{
  \institution{ETH Zürich} \city{Zurich} \country{Switzerland}
}

\author{Konstantin Lucny}
\orcid{0009-0006-5145-8662}
\affiliation{
  \institution{ETH Zürich} \city{Zurich} \country{Switzerland}
}

\author{Max Wierse}
\orcid{0009-0008-0324-2466}
\affiliation{
  \institution{ETH Zürich} \city{Zurich} \country{Switzerland}
}

\author{Samuel Gruetter}
\orcid{0000-0001-8369-9117}
\affiliation{
  \institution{ETH Zürich} \city{Zurich} \country{Switzerland}
}

\author{Timothy Roscoe}
\orcid{0000-0002-8298-1126}
\affiliation{
  \institution{ETH Zürich} \city{Zurich} \country{Switzerland}
}

\renewcommand{\shortauthors}{Fiedler, et al.}
\renewcommand{\shorttitle}{Sockeye: Bug-finding and proofs for platform
configurations and hardware \dots}

\begin{abstract}
  The ever increasing complexity of hardware platforms poses a challenge
  to systems programmers.
  Correctly programming a multitude of components, providing
  functionality \emph{and} security, is difficult: semantics of
  individual units are described in prose, underspecified, and prone to
  inaccuracies. Rigorous statements about platform security are often
  impossible.

  We are the first to address this problem for closed-source hardware
  platforms, by introducing a domain-specific language to formally
  describe hardware semantics based on hardware-reference manuals,
  assumptions about software behavior, and desired security properties.
  We demonstrate the practicality of our approach by creating
  machine-readable specifications for a diverse set of eight platforms
  from their reference manuals, and formally proving their
  (in-)security.
  In addition to security proofs about memory confidentiality and
  integrity, we discover a handful of documentation errors.
  Finally, our analysis also revealed a vulnerability on a real-world
  server chip, which was confirmed by the vendor to apply to a wide
  family of deployed network appliances.
  Our tooling offers system integrators a way of formally describing
  security properties for whole platforms, and the means to find
  counterexamples, or proving them correct.
\end{abstract}
 
\maketitle
\thispagestyle{empty}

\section{Introduction}
\label{sec:introduction}

Modern computing platforms, in particular \glspl{soc}, are enormously
complex artifacts, with programming manuals running to thousands of
PDF pages of informal human-readable text.  Much of the
software-visible complexity of an \gls{soc} resides in the security
features both of the core -- secure monitors,
privileged system management modes, \glspl{tee}, etc.\ -- and
the rest of the system -- on-chip protection units, interconnect
firewalls, etc.  Correctly configuring this hardware is of
critical importance, but this process is
fraught with errors, some of which can lie in the design of the hardware
itself.

\begin{figure}
  \includegraphics[width=0.9\columnwidth]{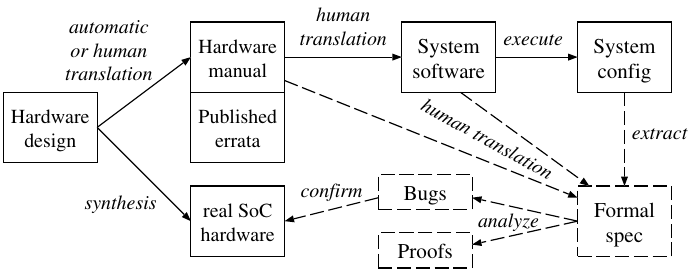}
  \caption{Traditional systems development (solid lines),
    plus \system analysis (dashed lines)}
  \label{fig:workflow}
\end{figure}

The solid lines in \cref{fig:workflow} show
the traditional process by which a (supposedly) secure configuration of
the hardware is achieved.
Vendors create documentation manually or semi-automatically
from the hardware design.  Based on this, developers write
system software (firmware or the OS) to configure the hardware.
Executing this software results in a \emph{configuration} of the
hardware: register contents, page tables, etc.\  intended to enforce
system-wide security properties.
Some architectural features can naturally be formalized as integrity and
confidentiality properties, such as \tz, Arm confidential
computing architecture, or RISC-V Worlds.
We can derive similar isolation properties for low-level hardware
configuration, which typically relies on static partitioning between
itself and the application-facing OSes like Linux.

Errors can, and do, occur at all stages of the process. First, hardware
may contain fundamental security flaws, such as AMD
Sinkclose~\cite{sinkclose}, which allowed a kernel to run arbitrary code
in \gls{smm}.
Second, either the hardware as described by the documentation
can deviate from the real implementation, or the
documentation can be ambiguous, failing to precisely
describe the hardware.  Software based on an accurate interpretation
of the documentation thus might still be insecure, either because
the documentation was wrong, or because the software developer chose
an incorrect (but plausible) interpretation of it.
Finally, software itself can be simply incorrect, generating a
configuration which fails to enforce security properties.

Existing bug-finding tools and formal verification might improve the
situation, but face the following challenges:
\begin{itemize}
\item The source code (RTL) of hardware is rarely publicly available,
thus tools that require RTL can rarely be used by the systems software
developers who might benefit.
\item Bug-finding tools~\cite{metal_PLDI02,hossain:date:2023,FormalFuzzer_ASPDAC24}
reveal individual bugs, but provide no guarantees about the
\emph{absence} of bugs, resulting in the endless whac-a-mole
game: there can always be more bugs.
\item Formal methods provide proofs of system properties,
but for non-experts it is hard to tell which bugs the
proof excludes, or are beyond scope of the model.
If a proof tool also supports bug-finding, users
can convince themselves that a bug is prevented by the proof by
inserting a bug and checking if the tool finds it, but unfortunately,
many popular proof tools, including Dafny, Verus, and Rocq, do not
provide counterexamples and therefore cannot be used for automatic bug
finding.
\end{itemize}

We address these challenges with \system, a tool for specifying and
analyzing systems whose RTL source is not available, albeit at the cost
of transcribing the manuals into a \system spec.
It is not intended to model \gls{isa} semantics, but to model those
parts that \gls{isa} specifications deliberately leave out,
namely the platform-specific interconnects and devices and
the semantics of their memory-mapped registers.

\system supports both bug
finding and proofs \emph{on top of the same system model}.
It addresses the limitation of bug-finding-only tools,
as proofs can exclude \emph{whole classes} of bugs.
\textbf{Moreover, ensuring that a specification is capable of expressing and
reproducing a buggy scenario provides evidence that the proofs indeed
exclude the kind of bugs we care about.}
While transcribing and analyzing entire manuals of several thousand
pages is not yet in reach, we can transcribe security-relevant
parts of a manual, and
each new section of the manual that is transcribed and
verified constitutes progress by excluding a
further class of bugs.  In contrast, running a bug finder
for longer or with more heuristics might find more bugs, but cannot
guarantee progress towards a secure system.

\system uses a single \gls{dsl} to express the behavior of the
hardware, the software that configures it, \emph{and} the desired
security properties.
It is just expressive enough for these three forms of specification,
while also supporting efficient \emph{bug and proof finding} based on
\gls{smt} solving.

We applied \system to eight different real-world platforms
for which we had access to documentation and, in most cases, hardware
boards.  On the bug-finding side, we identified numerous important
ambiguities in the documentation, found two critical documentation
errors, reproduced two known hardware bugs, and discovered a
\emph{previously-unknown critical vulnerability present in shipping
hardware}, which we have reported to the vendor
and has since been fixed by them.
On the proof side, we developed four proofs about security
properties for three of our modeled platforms.

We make the following contributions:
\begin{itemize}
  \item A formal \gls{dsl} for transcribing hardware manuals and
    describing intended security behavior
    (\cref{sec:overview,sec:language-design,sec:using-system}),
  \item a tool that translates \gls{dsl} specifications to formats suitable
    for automated verification tools to prove said properties or generate
    counterexamples (\cref{sec:backends}),
  \item specifications for a representative set of eight platforms, showing
    that our techniques can successfully be applied to real-world hardware, and
    can result in discovery of new bugs and vulnerabilities
    in real hardware as well as in proofs about
    security guarantees
    (\cref{sec:case-studies}), and
  \item an evaluation of \system and its \gls{dsl} (\cref{sec:evaluation}).
\end{itemize}

The \system tool and related specifications are publicly available on
GitLab\footnote{\url{https://gitlab.inf.ethz.ch/project-opensockeye/sockeye}}
under a permissive open-source license.
An artifact~\cite{sockeye-sosp26-zenodo} associated with this paper was
evaluated and is available at
\url{https://doi.org/10.5281/zenodo.21976745}.

\section{Overview by an example}
\label{sec:overview}

We introduce \system with a simplified example,
showing how we found a vulnerability in the \txone SoC,
and how it can prove absence of vulnerabilities (within the chosen
modeling precision).
We reported the bug to Marvell, the current vendor, who confirmed
our findings, and that it also affects their
current OCTEON processors, for which they issued a
fix. The bug is awaiting assignment of a CVE identifier.
We only print the most relevant snippets
of the example here, and describe the rest in English.
The simplified example is 274 LoC and is listed in full
in \cref{app:mini_thunderx1}.
Figures for how long \system takes to find the vulnerability
on our full model appear in \cref{tab:bug-runtimes}.

\subsection{Graph of components}

At the top level, a \system model is a graph of components.
\Cref{code:MiniThunderX1Toplevel} shows how we represent this graph structure
textually, as well as the corresponding graphical representation.

A \emph{module} specifies a blueprint for a component which can be
\emph{instantiated} multiple times as the child of another module.
Instances thus form a tree, but if an instance needs to access another
instance that is not its child, it can do so using the \lstinline{callee}
keyword.
When instantiating a module that has callees, one has to specify for each
callee an instance to which calls to that callee should be directed.
For example, \cref{line:dram_connection} specifies that calls that the
\lstinline{instance asc} inside \lstinline{MiniThunderX1} makes to its
\lstinline{callee dram} should be directed to the \lstinline{instance dram}
of \lstinline{MiniThunderX1}.

\begin{lstlisting}[numbers=left, float,
  caption = {Top-level structure of our example},
  captionpos = {b},
  label = {code:MiniThunderX1Toplevel}]
module Region {
  instance START: State<BitInt(64)>(0);
  instance END: State<BitInt(64)>(0);
  instance ATTR: State<BitInt(64)>(0);
  ...
}
module DRAM {
  instance storage: Array<BitInt(31), BitInt(64)>;`\label{line:DRAM_storage}`
  ...
}
module ASC { /* Address Space Controller */
  instance region0: Region;
  instance region1: Region;
  instance region2: Region;
  instance region3: Region;
  callee dram: DRAM;`%



\tikzset{every picture/.style={line width=0.75pt}} %

\begin{tikzpicture}[overlay,xshift=.5cm,yshift=.5cm,x=0.75pt,y=0.75pt,yscale=-1,xscale=1]
%

%
\draw  [fill={rgb, 255:red, 255; green, 255; blue, 255 }  ,fill opacity=1 ] (41,112) -- (121,112) -- (121,132) -- (41,132) -- cycle ;
%
\draw  [fill={rgb, 255:red, 255; green, 255; blue, 255 }  ,fill opacity=1 ] (37,116) -- (117,116) -- (117,136) -- (37,136) -- cycle ;
%
\draw  [fill={rgb, 255:red, 255; green, 255; blue, 255 }  ,fill opacity=1 ] (33,120) -- (113,120) -- (113,140) -- (33,140) -- cycle ;
%
\draw   (10,20) -- (140,20) -- (140,210) -- (10,210) -- cycle ;
%
\draw  [draw opacity=0][fill={rgb, 255:red, 255; green, 255; blue, 255 }  ,fill opacity=1 ] (16,11) -- (110,11) -- (110,31.16) -- (16,31.16) -- cycle ;
%
\draw   (20,90) -- (130,90) -- (130,150) -- (20,150) -- cycle ;
%
\draw  [fill={rgb, 255:red, 255; green, 255; blue, 255 }  ,fill opacity=1 ] (29,124) -- (109,124) -- (109,144) -- (29,144) -- cycle ;
%
\draw   (20,40) -- (130,40) -- (130,70) -- (20,70) -- cycle ;
%
\draw    (75,70) -- (75,88) ;
\draw [shift={(75,90)}, rotate = 270] [color={rgb, 255:red, 0; green, 0; blue, 0 }  ][line width=0.75]    (10.93,-3.29) .. controls (6.95,-1.4) and (3.31,-0.3) .. (0,0) .. controls (3.31,0.3) and (6.95,1.4) .. (10.93,3.29)   ;
%
\draw    (75,150) -- (75,168) ;
\draw [shift={(75,170)}, rotate = 270] [color={rgb, 255:red, 0; green, 0; blue, 0 }  ][line width=0.75]    (10.93,-3.29) .. controls (6.95,-1.4) and (3.31,-0.3) .. (0,0) .. controls (3.31,0.3) and (6.95,1.4) .. (10.93,3.29)   ;
%
\draw   (20,170) -- (130,170) -- (130,200) -- (20,200) -- cycle ;

%
\draw (18,13) node [anchor=north west][inner sep=0.75pt]   [align=left] {{\fontfamily{ptm}\selectfont MiniThunderX1}};
%
\draw (48,126) node [anchor=north west][inner sep=0.75pt]   [align=left] {{\fontfamily{ptm}\selectfont Region}};
%
\draw (59,93) node [anchor=north west][inner sep=0.75pt]   [align=left] {{\fontfamily{ptm}\selectfont ASC}};
%
\draw (75.58,55.5) node   [align=left] {{\fontfamily{ptm}\selectfont CPU}};
%
\draw (75,185) node   [align=left] {\begin{minipage}[lt]{33.3pt}\setlength\topsep{0pt}
\begin{center}
{\fontfamily{ptm}\selectfont DRAM}
\end{center}

\end{minipage}};


\end{tikzpicture}
 %
`
  ...
}
module CPU {
  callee asc: ASC;
  instance is_secure:
     State<Bool>(true);`\label{line:CPU_flag}`
  ...
}
module MiniThunderX1 {
  instance cpu: CPU;
  instance asc: ASC;
  instance dram: DRAM;
  asc.dram -> dram;`\label{line:dram_connection}`
  cpu.asc -> asc;
  ...
}
\end{lstlisting}

In real hardware, connections between components are wires transmitting
bits, above which one implements request/response protocols.
However, in \system we take a higher-level view and
model requests as function calls and replies as values returned by these
calls.

\subsection{Mutable state via primitive modules}

The only way to obtain mutable state in \system is via
\emph{primitive modules}.
For example, the \lstinline{State<Bool>(true)} module holds a boolean
value, initialized to \lstinline{true}. It has builtin getter/setter
methods, and is used to add a flag
to the \lstinline{CPU} module (\cref{line:CPU_flag}).
We also provide a primitive \lstinline{Array}
module that takes key and value type arguments, which is used to
model 16~GB of \lstinline{DRAM}, represented as an array
of $2^{31}$ $64$-bit integers (\cref{line:DRAM_storage}).

\subsection{The Address Space Controller}

The \emph{Address Space Controller} (ASC) of the \txone filters
CPU DRAM requests based on the \tz security policy.
\tz is a hardware-based mechanism providing two isolated execution
environments: the \emph{Secure} and \emph{Non-Secure worlds}, and
is orthogonal to the user/kernel/hypervisor/machine modes.

The ASC has four configurable regions, each with \lstinline{START} and \lstinline{END} fields,
plus an \lstinline{ATTR} field with one of three possible values:
\lstinline{ATTR_SEC} means that the region between \lstinline{START} and
\lstinline{END} can only be accessed if the CPU is currently in Secure mode,
\lstinline{ATTR_NONSEC} means that the range may also be accessed in Non-Secure
mode, and \lstinline{ATTR_EMPTY} means that the region is not used.
If the regions overlap, the hardware behavior is undefined.
\Cref{fig:region_config} shows a sample configuration where region 0 is used
for the Secure world, region 1 is used for the Non-Secure world,
and regions 2 and 3 are unused.

\begin{figure}
\tikzset{every picture/.style={line width=0.75pt}} %

\begin{tikzpicture}[x=0.75pt,y=0.75pt,yscale=-1,xscale=1]
\draw   (229,59.59) -- (340,59.59) -- (340,190) -- (229,190) -- cycle ;
\draw    (214,59.59) -- (340,59.59) ;
\draw    (214,139.59) -- (340,140) ;
\draw    (214,190) -- (340,190) ;

\draw (223,187) node [anchor=south east] [inner sep=0.75pt]  [font=\small] [align=left] {\texttt{region0.START = 0x0}};
\draw (223.09,143) node [anchor=north east] [inner sep=0.75pt]  [font=\small] [align=left] {\texttt{region0.END = 0xff\_ffff}};
\draw (223,137) node [anchor=south east] [inner sep=0.75pt]  [font=\small] [align=left] {\texttt{region1.START = 0x100\_0000}};
\draw (223,63.59) node [anchor=north east] [inner sep=0.75pt]  [font=\small] [align=left] {\texttt{region1.END = 0x3\_ffff\_ffff}};
\draw (111,12) node [anchor=north west][inner sep=0.75pt]  [font=\small] [align=left] {\texttt{region3.ATTR = ATTR\_EMPTY}};
\draw (111,32) node [anchor=north west][inner sep=0.75pt]  [font=\small] [align=left] {\texttt{region2.ATTR = ATTR\_EMPTY}};
\draw (241,82) node [anchor=north west][inner sep=0.75pt]  [font=\small] [align=left] {\texttt{region1.ATTR}\\\texttt{= ATTR\_NONSEC}};
\draw (241,147) node [anchor=north west][inner sep=0.75pt]  [font=\small] [align=left] {\texttt{region0.ATTR}\\\texttt{= ATTR\_SEC}};

\end{tikzpicture}
   \caption{Sample region configuration}
  \label{fig:region_config}
\end{figure}

\begin{lstlisting}[
  float,
  caption = {Record types for requests and responses},
  captionpos = {b},
  label = {code:request_response_types}]
type Request = {      `
\begin{tikzpicture}[overlay,yshift=3mm]
  \draw [line width=0.2mm, color={black!60}] (0, 0) -- (0, -2.15);
\end{tikzpicture}
` type PhysAddr = BitInt(48);
  is_write: Bool,      type Response = {
  is_secure: Bool,       ok: Bool,
  address: PhysAddr,     value: BitInt(64)
  value: BitInt(64)    };
};
\end{lstlisting}

\begin{lstlisting}[numbers=left,float,
  caption = {Memory-request handling function of the ASC},
  captionpos = {b},
  label = {code:request_handler}]
mut fn request(r: Request) -> Response {
  if is_region_config_addr(r.address) { `\label{line:is_region_config_addr_call}`
    asc_region_request(r)
  } else if is_allowed_dram_addr(r) { `\label{line:is_allowed_dram_addr_call}`
    dram_request(r)
  } else {
    { ok: false, value: any<BitInt(64)> }
  }
}
\end{lstlisting}

\subsection{Modeling memory requests}

To model memory requests the CPU sends to the ASC, we use the
\lstinline{Request} record type (shown in \cref{code:request_response_types}),
and for the responses from the ASC, we use \lstinline{Response}.
\Cref{code:request_handler} shows the request handler in
\lstinline{module ASC} that encodes (a simplification of) our understanding of
the \txone manual's description of how memory requests
are handled.
On \cref{line:is_region_config_addr_call}, the \lstinline{is_region_config_addr}
function (omitted here, but shown in \cref{app:mini_thunderx1})
determines if an address lies in the MMIO address range of the
region-configuring registers,
while the \lstinline{config_write} and \lstinline{config_read} helper functions
read or update the \lstinline{START}, \lstinline{END} or \lstinline{ATTR}
fields of region 0, 1, 2, or 3, depending on values in the
request record~\lstinline{r}.
On \cref{line:is_allowed_dram_addr_call}, \lstinline{is_allowed_dram_addr}
checks if the request is allowed,
given the \lstinline{is_secure} flag in the request that indicates
whether the request comes from the Secure or Non-Secure world, and given the
current configuration of the ASC regions.
If it is, the request is forwarded to \lstinline{dram}.

\begin{lstlisting}[numbers=left, float, captionpos={b},
  caption = {Example encoding the property that, in two computation steps,
  Non-Secure code cannot modify Secure memory},
  label = {code:test_secure_area_unchanged},
]
module Main {
  instance miniTX1: MiniThunderX1;
  ...
  mut fn test_secure_area_unchanged() {
    setup_regions();
    miniTX1.cpu.is_secure.set(false);`\label{line:set_non_secure}`
    let orig_mem = miniTX1.dram.storage.get();`\label{line:dram_snapshot}`
    miniTX1.step();
    miniTX1.step();
    let new_mem = miniTX1.dram.storage.get();`\label{line:second_dram_snapshot}`
    let test_addr = any<BitInt(31)>;
    assume(test_addr <= 0x1f_ffffu31);
    assert(orig_mem[test_addr]==new_mem[test_addr])
  }
}
\end{lstlisting}

\subsection{Modeling a security property}

Now, we can precisely and unambiguously state
what we mean by ``the address space controller prevents Non-Secure contexts
from writing to memory that belongs to Secure contexts,'' as shown in
\cref{code:test_secure_area_unchanged}:
We create a \lstinline{Main} module containing a
\lstinline{test_secure_area_unchanged} function that encodes the following
\emph{scenario}:

First, \lstinline{setup_regions()} writes the values in
\cref{fig:region_config} to the ASC region-configuring registers.
This code (omitted here, but shown in \cref{app:mini_thunderx1}),
would be in C or assembly on a real system, but here
we express it in the \system language.%

Next, we specify that the CPU is currently executing in the Non-Secure
world (\cref{line:set_non_secure}),
and take a snapshot of the current DRAM contents (\cref{line:dram_snapshot}).
We then perform two computation steps of \lstinline{MiniThunderX1}.
Two is an arbitrary choice but, as we shall see, is sufficient in this case
to detect a vulnerability.  In practice a search for bugs would try
as many steps as the solver can deal with within reasonable time.

In our simplified example, one computation step of the SoC is just one
computation step of the CPU, but in the more complete \txone model,
the step function \emph{nondeterministically picks} one component that makes a step,
(e.g. CPU, ZIP hardware accelerator, network interface, etc.).
The CPU step function is given in \cref{code:cpu_step}.
It uses the keyword \lstinline{any} to populate a request
with nondeterministic values, except for the \lstinline{is_secure}
field, which is assigned the CPU flag that determines if it is
executing in Secure or Non-Secure world.

\begin{lstlisting}[
  float,
  captionpos = {b},
  caption = {Step function of the CPU},
  label = {code:cpu_step},
]
mut fn step() {
  let r: Request = {
    is_write: any<Bool>,
    is_secure: is_secure.get(),
    address: any<PhysAddr>,
    value: any<BitInt(64)>
  };
  printf("CPU: request is {r}\n");
  let ignored_reply = asc.request(r); ()
}
\end{lstlisting}

Finally, on \cref{line:second_dram_snapshot} in
\cref{code:test_secure_area_unchanged},
we again snapshot the DRAM,
pick an arbitrary \lstinline{test_addr} that we \lstinline{assume} to lie
within the secure memory range,
and \lstinline{assert} that the snapshots from before and after running the
two steps agree on the value present at the chosen address,
which amounts to asserting that all the values within the secure memory
range remained unchanged.
This property, which we expressed using the
nondeterministic \lstinline{any} and the \lstinline{assume} and
\lstinline{assert} keywords, could also be expressed using more mathematical
notation:
\[ \forall a \le \texttt{0x1f\_ffff}, \texttt{orig\_mem}[a] = \texttt{new\_mem}[a] \]
However, we deliberately omitted $\forall$ and $\exists$ quantifiers from the
\system language to make the analysis tractable for solvers.

\subsection{Automatically finding security violations}
\label{sec:finding_secure_area_changed_vuln}

\system has a backend emitting Rosette \cite{Rosette_PLDI14},
a language for SMT-based symbolic execution.
Rosette starts symbolic execution at \texttt{\small{test\-\_secure\-\_area\-\_unchanged}}
and explores all possible branches, accumulating a single SMT formula.
Whenever it encounters a nondeterministic \lstinline{any<T>} choice for some
type \lstinline{T}, it declares a symbolic variable of type \lstinline{T}.
At the end, it sends a query to the Z3 solver for an assignment of concrete values for
all symbolic variables so that all \lstinline{assume} statements are
\lstinline{true}, and at least one \lstinline{assert} statement is
\lstinline{false}.

\begin{lstlisting}[keywords={},
  float,
  captionpos = {b},
  caption = {Printf output of the attack (the output of the initial
    \lstinline{setup_regions()} call is omitted)},
  label = {code:attack_trace},
  breaklines=true,
  postbreak=\mbox{\textcolor{gray}{$\hookrightarrow$}\space},
]
CPU: request is { address: 0x8000_0000_0070u48, is_secure: false, is_write: true, value: 1 }
ASC: Setting region3.ATTR to 1
CPU: request is { address: 0, is_secure: false, is_write: true, value: 0x48ad_c33c_fdc9_99d4u64 }
DRAM: Storing 0x48ad_c33c_fdc9_99d4u64 to 0
\end{lstlisting}

On the above example, Z3 does find such an assignment, and \system then runs
its interpreter on the original program, substituting the values of the
assignment wherever the program uses the \lstinline{any} keyword,
in order to find out which of the assertions failed.
\Cref{code:attack_trace} shows \system's output after an execution that
violated the \lstinline{secure_area_unchanged} property.

\subsection{Understanding the vulnerability}

Note the first CPU memory request
is to \lstinline{0x8000_0000_0070}, the \lstinline{ATTR} configuration register of region 3,
and that (on the next line) the value written to it is 1,
which means that the Non-Secure world can access region 3.
The second request from the CPU is at address \lstinline{0}, i.e.\ in
the secure DRAM region as well as in region 3,
and writes some garbage value whose only restriction is
that it is different from the original value.

So, we see that the problem is that the registers which control which DRAM
regions can be accessed from the Non-Secure world are writable by the Non-Secure
world, and therefore the Non-Secure world can also access DRAM regions that are
supposed to be only accessible from the Secure world.

\subsection{Fixing the vulnerability}
\label{sec:fixing_secure_area_changed_vuln}

According to both the \txone manual and our subsequent testing, the
region-configuration registers of the ASC are indeed
writable from the Non-Secure world, but for the sake of exploration,
let us modify our \system model so that the security property now holds,
which we can achieve by inserting ``\lstinline{&& r.is_secure}''
in the check
on \cref{line:is_region_config_addr_call} in \cref{code:request_handler}.

If we now rerun \system, Z3 finds no satisfying assignment, reporting \lstinline{unsat} instead,
which means that within two steps of \lstinline{MiniThunderX1}, the security property
cannot be violated.

\begin{lstlisting}[
  float,
  captionpos = {b},
  caption = {Induction proof},
  label = {code:example_induction_proof}
]
  mut fn base_case() {
    setup_regions();
    miniTX1.cpu.is_secure.set(false);
    assert(invariant());
  }
  mut fn inductive_step() {
    miniTX1.havoc();
    assume(invariant());
    miniTX1.step();
    assert(invariant());
  }
  mut fn invariant_is_useful() {
    miniTX1.havoc();
    assume(invariant());
    let orig_mem = miniTX1.dram.storage.get();
    miniTX1.step();
    let new_mem = miniTX1.dram.storage.get();
    let test_addr = any<BitInt(31)>;
    assume(test_addr <= 0x1f_ffffu31);
    assert(orig_mem[test_addr] == new_mem[test_addr])
  }
\end{lstlisting}

\subsection{Proving absence of vulnerabilities}
\label{sec:example_induction_proofs}

No matter how many invocations of \lstinline{miniTX1.step()} we insert in
\cref{code:test_secure_area_unchanged}, the best guarantee that we can
obtain is that the security property holds for a
\emph{bounded} number of steps.
To prove that it holds for an \emph{unbounded} number of steps,
we need a different approach, based on a \emph{proof by induction} over the
number of steps.
An induction proof needs an \lstinline{invariant} that holds initially (checked
by the \lstinline{base_case} scenario in \cref{code:example_induction_proof})
and that is also preserved by each step
(checked by the \lstinline{inductive_step} scenario, which first also uses
\lstinline{miniTX1.havoc()} to make sure the system is in a completely arbitrary
state, instead of starting from the reset values).
In our example, we choose an invariant that checks that all the four regions
either have their \lstinline{ATTR}
set to \lstinline{ATTR_EMPTY} or \lstinline{ATTR_SEC},
or have a range that is disjoint from the secure range to be protected.
The full code for this \lstinline{invariant} is given in
\cref{app:mini_thunderx1}.
If no assertion violations are found in \lstinline{base_case} or
\lstinline{inductive_step}, we know that \lstinline{invariant} always holds,
and using a third scenario, \lstinline{invariant_is_useful}, we can check
that our invariant implies the desired security property.

\section{Language Design}
\label{sec:language-design}

\setlength{\grammarparsep}{4pt plus 1pt minus 1pt}
\renewcommand{\grammarlabel}[2]{\textit{#1} \hfill #2}
\renewcommand{\syntleft}{\normalfont\itshape}
\renewcommand{\syntright}{}
\renewcommand{\ulitleft}{\normalfont\bfseries\ttfamily\frenchspacing}
\renewcommand{\ulitright}{}
\begin{figure}
  \begin{grammar}
    \footnotesize
  <program> ::= <top-level-defs>

  <top-level-def> ::= <type-alias>
    \alt <module>
    \alt <interface>
    \alt <constant>
    \alt <fn-def>

  <type-alias> ::= "type" <ident> "=" <type>

  <module> ::= "module" <ident> "{" <module-members> "}"

  <interface> ::= "interface" <ident> "{" <fn-signatures> "}"

  <constant> ::= "const" <ident> ":" <type> "=" <expr>

  <fn-def> ::= <fn-signature> "{" <expr> "}"

  <module-member> ::= "instance" <ident> ":" <module-name>
    \alt "callee" <ident> ":" (<module-name> | <interface-name>)
    \alt <constant>
    \alt <fn-def>
    \alt <caller-callee-wiring>

  <caller-callee-wiring> ::= <qualified-ident> "->" <qualified-ident>

  <fn-signature> ::= <fn-effect> "fn" <ident> "(" <typed-params> ")" <rtype>

  <fn-effect> ::= "mut" | <empty>

  <rtype> ::= "->" <type> | <empty>

  <type> ::= "()" | "Bool" | "BitInt(" <n> ")"
    \alt <variant> "|" <variant> "|" \dots
    \alt "Vec(" <type> "," <n> ")"
    \alt "{" <record-fields> "}"

  <expr> ::= <n> | "0x" <n> | "0b" <n> | <n> <op> <n> | \dots
    \alt "if" <expr> "{" <expr> "} else {" <expr> "}"
    \alt "let" <v> "=" <expr> ";" <expr>
    \alt <fn-name> "(" <params> ")"
    \alt \dots
\end{grammar}
  \caption{BNF grammar of the \system language}
  \label{fig:system-grammar}
\end{figure}

The \system language is a \gls{dsl} that models a hardware platform's
functionality, and state.
The \system compiler analyzes and transforms a specification in this \gls{dsl}
to a form suitable for automatic verification.
\Cref{fig:system-grammar} shows an excerpt of the grammar underlying
the \system \gls{dsl}.

A program is a collection of top-level definitions. Module instances form a
graph of components, which define the overall structure of a \system model.
A module groups state and behavior of \system components of any size, from
individual registers to entire platforms. It is a \emph{blueprint} which
can be \emph{instantiated} any number of times, with each instance providing a
unique copy of its state and functionality. Modules contain \emph{instances} of other
modules, and \emph{callees}, which allow interaction with non-child instances.
Functions describe the behaviors module instances. A module can access and
modify the state and behavior of its child instances, and any connected callees
via the defined interface. Callees instances must be wired up by an ancestor
module, and must resolve to a valid instance. For modeling purposes, an instance
models the ``consists-of'' relationship of a hardware component, while a callee
is closer to a ``depends-on'' or ``requires''.

An interface is a named set of function signatures. To implement an
interface, it suffices that a module implements each function
signature. Interfaces allow components to define
boundaries or dependencies in an implementation-agnostic way:
an SMMU module, say, might define an interface for interactions
with the \gls{soc}-specific downstream interconnect, without relying
on platform specifics.

\system provides two built-in \emph{primitive modules} which model
stateful components, from single-bit registers to terabytes of memory.
\lstinline{State<T>(expr)} represents a single, mutable value
of type \lstinline{T} with initial value \lstinline{expr}.\footnote{Note that
\lstinline{any} is a valid value for \lstinline{expr}.}, while
\lstinline{Array<K, V>} provides a vector of individual
\lstinline{V}s, indexed by \lstinline{K}s, and allows access to both
the entire collection, individual elements, or sub-ranges. \system supports efficient sparse array representation
without individual element allocation.

Function bodies are expressions, with support for
non-determinism and underspecification. They support sequential composition and
branching, but no loops.  This condition does not prevent us
from expressing complex security properties and proofs in \system (see
\cref{sec:security_properties}).

All \system expressions are statically and strongly typed. The type system's
primitive types include fixed-width integers, unbounded integers, enumerations.
Above this, it provides fixed-length vectors and records with named
fields. The language supports local type inference and bidirectional
type checking~\cite{BidirTC_POPL98,BidirTC_CSUR21}. Bit-widths of integer expressions
are always known statically, and precisely tracked.

Underspecification is supported via the \lstinline{any<T>} expression, which
non-deterministically chooses an arbitrary, valid value of type T. This proves a
useful tool when constructing a specification from reference manuals:
technical ambiguities and forward references can safely be over-approximated by
allowing \emph{any} valid value in a certain spot.

Using \lstinline{assume(prop)} and \lstinline{assert(prop)}, where
\lstinline{prop} is a boolean expression, we can encode verification assumptions
and conditions. During verification, these expressions denote restrictions (assumptions)
and proof obligations (assertions), which the verifier aims to (dis-)prove.
The \system language is designed to give precise errors in cases of assertion
violations: when \system determines that an assertion is falsifiable it can
provide a trace of all required non-deterministic choices made that lead to the
assertion violation.

Decoding thousands of lines of call traces can be tedious, so \system provides a
debugging aid via \lstinline{printf(s)}, where \lstinline{s} denotes a format
string to emit when called. During verification, these
statements are no-ops, however during reconstruction they can be used to
highlight particular choices made by the underlying solver, for example which
top-level step was taken, what address or value was chosen to be written, etc.

Our \gls{dsl} is a ``sweet spot'' between a hardware description language
and a conventional programming language.  We do not try to capture
low-level hardware details such as cycle-accurate behavior or \gls{rtl}, but
we do care about bit-level hardware representation in certain cases.

Module \lstinline{Main} is the root of the instance tree, and
is instantiated exactly once. It may not contain callees, and
all callees of its instances must be wired to valid targets. Top-level
functions defined in \lstinline{Main} are entry points for \system
\emph{scenarios}, and form a single test-case, bug-finding attempt, or proof.

\section{Using \system}
\label{sec:using-system}

The first step in modeling a platform consists of identifying the
different modules and modeling their state. Usually, one module
denotes the root platform, which contains instances of modules
that represent individual hardware components, and ensures everything
is wired up suitably.
Next, one models their behavior using methods like e.g. in
\cref{code:request_handler}.
Typically, one also adds a top-level \lstinline{step()} function that
nondeterministically picks one component that takes a step.
The step granularity is a design tradeoff between the model complexity
and the degree of possible interleavings one wants to model.
Next, one models the security properties of interest
(described further in \cref{sec:security_properties}).
Next, it is best to write scenarios that test the whether the
properties hold after 1, 2, or 3 steps
(e.g. \cref{code:test_secure_area_unchanged}),
which often already reveals bugs in the system or (more likely)
in its modeling.
Finally, to turn the tests that explore a \emph{bounded} number of
steps into \emph{exhaustive} proofs, one uses induction proofs as
explained in \cref{sec:example_induction_proofs}.
Thanks to the \gls{smt}-based verification, the proofs are fully
automatic, and do not require any proof code besides the invariants
and the desired properties.
Usually, the invariants simply state what parts of the system
configuration must remain unchanged, which is much simpler than the
notoriously-hard-to-find invariants typically needed in proofs about
algorithms.

As is common in formal modelling, our results rely on additional
assumptions that we make about the hardware, its documentation, our
transcription, and the \system toolchain. If one of the following
assumptions does not hold, a bug or proof found using \system may not
hold in the real world:
\begin{itemize}
  \item The hardware reference manual specifies the actual behavior of
    the hardware.
  \item Our translation of the English text in the hardware reference
    manual into \system is accurate.
  \item The parts that we did not model in detail are over-approximated
    in a sound way, i.e.\ each actual behavior of the hardware is also a
    possible behavior of our spec.
  \item The omitted parts are not security-critical.
  \item The \system compiler correctly encodes the programs in the
    backends' input language.
  \item\label{asm:backends-correct} The backends, Rosette, CBMC, and the SMT solvers they are using,
    are correct.
  \item All the compilers used to compile our tool and the backends, and
    the surrounding OS that we use to invoke \system, are correct.
\end{itemize}

\begin{figure}

\tikzset{every picture/.style={line width=0.75pt}} %

\begin{tikzpicture}[x=0.75pt,y=0.75pt,yscale=-1,xscale=1]
\draw    (184.84,39.54) -- (234.08,39.42) ;
\draw [shift={(236.08,39.42)}, rotate = 179.86] [color={rgb, 255:red, 0; green, 0; blue, 0 }  ][line width=0.75]    (10.93,-3.29) .. controls (6.95,-1.4) and (3.31,-0.3) .. (0,0) .. controls (3.31,0.3) and (6.95,1.4) .. (10.93,3.29)   ;
\draw    (116.5,70.9) -- (197.63,105.42) ;
\draw [shift={(199.47,106.2)}, rotate = 203.05] [color={rgb, 255:red, 0; green, 0; blue, 0 }  ][line width=0.75]    (10.93,-3.29) .. controls (6.95,-1.4) and (3.31,-0.3) .. (0,0) .. controls (3.31,0.3) and (6.95,1.4) .. (10.93,3.29)   ;
\draw    (116.5,70.9) -- (77.95,105.86) ;
\draw [shift={(76.47,107.2)}, rotate = 317.8] [color={rgb, 255:red, 0; green, 0; blue, 0 }  ][line width=0.75]    (10.93,-3.29) .. controls (6.95,-1.4) and (3.31,-0.3) .. (0,0) .. controls (3.31,0.3) and (6.95,1.4) .. (10.93,3.29)   ;
\draw    (201.13,142.2) -- (103.7,179.85) ;
\draw [shift={(101.83,180.57)}, rotate = 338.87] [color={rgb, 255:red, 0; green, 0; blue, 0 }  ][line width=0.75]    (10.93,-3.29) .. controls (6.95,-1.4) and (3.31,-0.3) .. (0,0) .. controls (3.31,0.3) and (6.95,1.4) .. (10.93,3.29)   ;
\draw    (201.13,142.2) -- (226.97,178.57) ;
\draw [shift={(228.13,180.2)}, rotate = 234.61] [color={rgb, 255:red, 0; green, 0; blue, 0 }  ][line width=0.75]    (10.93,-3.29) .. controls (6.95,-1.4) and (3.31,-0.3) .. (0,0) .. controls (3.31,0.3) and (6.95,1.4) .. (10.93,3.29)   ;
\draw    (93.03,218.59) -- (132.63,248.99) ;
\draw [shift={(134.22,250.21)}, rotate = 217.52] [color={rgb, 255:red, 0; green, 0; blue, 0 }  ][line width=0.75]    (10.93,-3.29) .. controls (6.95,-1.4) and (3.31,-0.3) .. (0,0) .. controls (3.31,0.3) and (6.95,1.4) .. (10.93,3.29)   ;
\draw    (93.03,218.21) -- (48.02,249.08) ;
\draw [shift={(46.37,250.21)}, rotate = 325.56] [color={rgb, 255:red, 0; green, 0; blue, 0 }  ][line width=0.75]    (10.93,-3.29) .. controls (6.95,-1.4) and (3.31,-0.3) .. (0,0) .. controls (3.31,0.3) and (6.95,1.4) .. (10.93,3.29)   ;

\draw (241,27.67) node [anchor=north west][inner sep=0.75pt]   [align=left] {\begin{minipage}[lt]{44.91pt}\setlength\topsep{0pt}
\begin{center}
{\fontfamily{ptm}\selectfont optimize}\\{\fontfamily{ptm}\selectfont and re-run}
\end{center}

\end{minipage}};
\draw (65.67,5) node [anchor=north west][inner sep=0.75pt]   [align=left] {\begin{minipage}[lt]{74.75pt}\setlength\topsep{0pt}
\begin{center}
{\fontfamily{ptm}\selectfont \textit{\textbf{Sockeye scenario}}}\\{\fontfamily{ptm}\selectfont hardware spec +}\\{\fontfamily{ptm}\selectfont platform config +}\\{\fontfamily{ptm}\selectfont security property}
\end{center}

\end{minipage}};
\draw (177.7,110) node [anchor=north west][inner sep=0.75pt]   [align=left] {\begin{minipage}[lt]{33.84pt}\setlength\topsep{0pt}
\begin{center}
{\fontfamily{ptm}\selectfont analyze}\\{\fontfamily{ptm}\selectfont trace}
\end{center}

\end{minipage}};
\draw (20,110) node [anchor=north west][inner sep=0.75pt]   [align=left] {\begin{minipage}[lt]{66.45pt}\setlength\topsep{0pt}
\begin{center}
{\fontfamily{ptm}\selectfont platform config}\\{\fontfamily{ptm}\selectfont is secure wrt.}\\{\fontfamily{ptm}\selectfont hardware spec}
\end{center}

\end{minipage}};
\draw (53.5,186.08) node [anchor=north west][inner sep=0.75pt]   [align=left] {\begin{minipage}[lt]{58.48pt}\setlength\topsep{0pt}
\begin{center}
{\fontfamily{ptm}\selectfont test bug on}\\{\fontfamily{ptm}\selectfont real hardware}
\end{center}

\end{minipage}};
\draw (192.67,186.08) node [anchor=north west][inner sep=0.75pt]   [align=left] {\begin{minipage}[lt]{60.19pt}\setlength\topsep{0pt}
\begin{center}
{\fontfamily{ptm}\selectfont change config}\\{\fontfamily{ptm}\selectfont and re-run}
\end{center}

\end{minipage}};
\draw (15.9,254.63) node [anchor=north west][inner sep=0.75pt]   [align=left] {\begin{minipage}[lt]{40.65pt}\setlength\topsep{0pt}
\begin{center}
{\fontfamily{ptm}\selectfont hardware}\\{\fontfamily{ptm}\selectfont is unsafe}
\end{center}

\end{minipage}};
\draw (95.53,254.63) node [anchor=north west][inner sep=0.75pt]   [align=left] {\begin{minipage}[lt]{63.32pt}\setlength\topsep{0pt}
\begin{center}
{\fontfamily{ptm}\selectfont documentation}\\{\fontfamily{ptm}\selectfont is unfaithful}
\end{center}

\end{minipage}};
\draw (1,79.5) node [anchor=north west][inner sep=0.75pt]  [font=\small] [align=left] {{\fontfamily{ptm}\selectfont {\small \textit{assertions verified}}}};
\draw (173,79.5) node [anchor=north west][inner sep=0.75pt]  [font=\small] [align=left] {{\small {\fontfamily{ptm}\selectfont \textit{counter-example}}}};
\draw (191,20.5) node [anchor=north west][inner sep=0.75pt]  [font=\small] [align=left] {{\small {\fontfamily{ptm}\selectfont \textit{timeout}}}};
\draw (216,147.5) node [anchor=north west][inner sep=0.75pt]  [font=\small] [align=left] {{\small {\fontfamily{ptm}\selectfont \textit{config bug}}}};
\draw  [draw opacity=0]  (136.8,161.1) -- (183.8,161.1) -- (183.8,183.1) -- (136.8,183.1) -- cycle  ;
\draw (139.8,165.1) node [anchor=north west][inner sep=0.75pt]  [font=\small] [align=left] {{\small {\fontfamily{ptm}\selectfont \textit{HW bug}}}};

\end{tikzpicture}

  \caption{Bug-finding workflow}
  \label{fig:flowchart}
\end{figure}

\subsection{Bug-finding workflow}

As shown in \cref{fig:flowchart},
given a specification of a hardware platform, together with a
set of required security properties, \system can determine if a configuration of
the platform \emph{as specified} satisfies the properties.

If this is not the case, \system generates a sequence of operations on
the platform, together with their results, which -- according to the
specification -- violate a security property.
Now we must analyze the exploit to determine whether the underlying
hardware is inherently unsafe, or whether another configuration would prevent it.
If we decide to try a different configuration (which corresponds to a software
modification), we retry our analysis with a new configuration.

However, if we conclude that no configuration can prevent this
exploit, we can check our exploit on the real hardware to distinguish the two possible
situations: a documentation bug (the hardware is correct, but
the spec, and by extension the manual is incorrect), or hardware bug
(the spec accurately describes unsafe hardware behavior).

In the case of a documentation bug, the spec can be updated to reflect
the observed behavior of the real hardware, the bug noted, and the
process repeated with the new spec. Otherwise, we have found a
hardware bug.

There remains one unresolved possibility, which is that \system
determines that the platform spec and a given configuration do satisfy
the required security properties, but the manual is incorrect and the
hardware implementation fails in this regard.   This case is the limit
of \system's ability to find problems, since it works on translations
of documentation.  The solution here is to start instead from the
hardware, using techniques such as fuzzing or to use formal
verification to show that the hardware design refines the \system
specification derived from the manual.

As an example, the vulnerability described in
\cref{sec:finding_secure_area_changed_vuln} is categorized as
``unsafe hardware, faithful documentation''.
With the hardware fix described in \cref{sec:fixing_secure_area_changed_vuln}
the \lstinline{MiniThunderX1} now falls into the category ``safe
hardware, faithful documentation''.
If this fix were applied to hardware, but the software incorrectly
configured the ASC, \system would still find a vulnerability and it
would be classified as ``safe hardware, insecure configuration''.

\subsection{Modelling security properties}
\label{sec:security_properties}

Using the primitives provided by the \system \gls{dsl}, users can define custom
properties. Typical properties of interest include confidentiality and
integrity statements, non-interference proofs, and inductive invariants.
In the following, we describe how to express different kinds of
properties in \system,
and corresponding code snippets can be found in \cref{app:sample_properties}.

\subsubsection{Directly testing the output of an access-control function}
\label{sec:test-access-control-fun}

The simplest, lowest-level check we can do is to set up the system,
perform a few arbitrary computation steps on it (using the \lstinline{any}
keyword to pick the actions and their arguments for each step).
Each of these steps, or potentially a combination of several steps,
might be able to change the active access control policies in undesired ways.
After performing these steps, we can check whether the return value of some
access-control function (e.g.\ inside a memory protection unit)
still correctly classifies requests as allowed and disallowed.

This approach reasons in terms of individual access-control mechanisms, and
might miss the bigger picture. Therefore, we also use the following
higher-level approaches to specify security properties.

\subsubsection{Comparing initial and final state snapshots}
\label{sec:compare-initial-final-snapshots}

Many properties can be checked using the approach shown in
\cref{code:test_secure_area_unchanged}:
take a snapshot of the initial state,
run some modifications,
take another snapshot,
and assert that they only differ in allowed ways,
e.g.\ that a certain memory area remained unchanged,
or that only allowed values were written.

\subsubsection{Specifying that an area cannot be read using a monitor}
\label{sec:read-write-monitor}

However, the approach used in the example in \cref{sec:overview} that compares
two snapshots of the memory can only detect writes, but not reads.
A simple way to detect reads as well is to interpose a monitor between the ASC and the DRAM.
Note that by ``monitor'', we mean a construct only used for the analysis,
not intended to represent any hardware.
In the \txone example, the monitor would wrap the DRAM, expose the exact same
interface as the DRAM, and intercept and forward all calls.
Whenever an undesired access (read or write) occurs, the monitor sets a boolean
flag, and at the end, we can assert that the flag was not set.

\subsubsection{Expressing integrity as preservation of an invariant}
\label{sec:integrity-as-invariant}

Often, integrity and absence of privilege escalation can be shown by showing
that an invariant about the access-control configuration registers is preserved.
In order to show absence of privilege escalation, it does not even matter what
exactly the invariant is, as long as it is not just the trivial \lstinline{true},
because an attacker who gained privileged access would be able to choose
a step for the system to take exactly tailored so that it violates the invariant,
so if the invariant cannot be violated, we also know that there is no possibility
for a privilege escalation.

\subsubsection{Expressing noninterference/information-flow properties using two copies of the \gls{soc}}
\label{sec:two-copies-noninterference}

In order check information-flow properties, we can distinguish
low-confidentiality and high-confidentiality data, and set up a
``proof bench'' where we instantiate two copies of the \gls{soc} to be analyzed,
in such a way that their initial states are what's called \emph{low-equivalent},
i.e.\ they agree on all values of low confidentiality, but might differ on
values of high confidentiality.
After running both copies for a few steps, making sure that both copies use
the same non-deterministic choices, we can check if they are still
low-equivalent, and if they are not, we know that the difference in
low-confidentiality data must come from leaks of the differing
high-confidentiality data.

\paragraph{Working without quantifiers}
Note that at first sight, it seems that \system cannot express
low-equivalence between two copies of the
\gls{soc}, because that would require a $\forall$ to quantify over all
low-confidentiality addresses.
But fortunately, as shown in \cref{code:test_secure_area_unchanged},
in order to \emph{assert} that a sub-range of a vector equals a
sub-range of another vector, we can emulate the $\forall$ quantifier
with a nondeterministic \lstinline{any} choice.
But it becomes harder when we want to \emph{assume} such a fact.
For each usage of the \lstinline{any} construct, the solver just tries to pick a
value that satisfies all assumptions and violates at least one assertion,
so if we used \lstinline{any} to pick a \lstinline{test_addr} nondeterministically,
we would only assume that the two vectors agree
\emph{at one specific address}, instead of assuming that they agree
\emph{at all addresses}.
The workaround for this limitation is to write separate
\lstinline{assume_low_equiv} and \lstinline{assert_low_equiv} functions.
In the \lstinline{assume_low_equiv} function,
in order to assume that a memory region contains certain values,
we simply copy the desired values to
the intended destination.%
And in the \lstinline{assert_low_equiv} function, we use \lstinline{any}
to compare the vectors at an adversarially-chosen address.

\subsubsection{Expressing integrity and non-interference as equivalence with an air-gapped system}
\label{sec:integrity-with-air-gapped-system}

There is an alternative way of expressing integrity and non-interference,
used e.g.\ in \cite{EnclaveIsol_CCS24},
which also involves running two systems in parallel, but contrary to the previous
approach, where we used two copies of the \emph{same} system,
we now use two different copies:
\begin{enumerate}
\item The more realistic model (which we could call ``implementation''
  in this context, though still a model of the actual
  implementation) uses intertwined state, e.g.\ RAM that contains a
  mix of different processes' data and modeled as a map from
  physical addresses to bytes.
\item The air-gapped model (which we could call ``specification'')
  uses very-obviously-separate state, e.g.\ the virtual memory of each
  process is modeled as a separate map from virtual addresses to
  bytes.
\end{enumerate}
Then, we can use \lstinline{any} to generate a schedule of which process runs
when, as well as to model how each process reacts to different inputs.
If we then run the two models in parallel, on the same initial data and same input data and same schedule, and assert that they behave the same, we can
detect a large class of isolation bugs:
For instance, if in the intertwined-state model, a process can access data of
another process because of an out-of-bounds read, the same access is unlikely
to succeed on the air-gapped model, so the differing behavior of the two models
points us to the bug.

\section{Backends}
\label{sec:backends}

The restricted nature of the \system language makes it possible to support a
multitude of backends. Specifications consist of sequential actions with
branching, which are supported by all tools in the verification space.
We disallowed loops in \system in hopes to keep solving times short.
Backends may support different sets of \system features.
\system has an interpreter backend and three verification backends:

\paragraph{Interpreter} The ``eval'' backend can be used to execute test
scenarios with concrete inputs.
It also has limited support for nondeterministic values stemming from the
\lstinline{any} keyword, by propagating unknown values using a simple form
of abstract interpretation~\cite{AbsInt_POPL77}.
It can also be used to reconstruct violating traces found by the Rosette
backend described below.

\paragraph{Direct translation to SMT} We initially encoded \system
scenarios as SMT queries directly, without going through an
intermediate language or tool.
While this encoding was useful for us to get started, this backend is
considerably less complete and performant than Rosette or CBMC.

\paragraph{Rosette} A Racket-based programming language which supports
symbolic execution~\cite{Rosette_Onward13,Rosette_PLDI14}.
Rosette has successfully been used as a verification backend
in Pensieve~\cite{Pensieve_ISCA23}. Under the hood, Rosette relies on the
Z3 SMT solver. It employs sophisticated techniques to generate the SMT
query, and outperforms our initial home-grown SMT encoding.

\paragraph{CBMC} A bounded model checker for C/C++
programs~\cite{kroening:cbmc:2014}, which can verify a C
program's memory safety, absence of undefined behavior, and even custom
assertions. C programs generated from \system specifications are memory safe
out-of-the-box, hence we disable most CBMC checks and only search for
violations of explicit assertions.

We design our verification procedure to produce explicit counterexamples in
cases of assertion failure. The advantage is that a concrete counterexample
can easily be checked against real-world hardware: set up the hardware with the
right initial values, and feed others in as necessary. Thus, we can quickly
determine whether behavior deemed insecure by the hardware manual is actually
exhibited by the hardware or not.

Most verification tools, including Rosette and Z3, support symbolic
representations for \system vectors and arrays, avoiding the need to allocate
billions or even trillions of elements. However, in cases where this is not
supported (as is the case for CBMC), we use a sparse representation which
represents an array as list of modifications with a configurable capacity of
stores. Until the number of stores reaches capacity, this representation
efficiently models (conceptually) huge arrays.

Supporting multiple backends allows us to use a portfolio-based approach
for verification and bug-finding. While Rosette usually outperforms
CBMC, this is not true in general as seen in \cref{sec:evaluation}.
This also presents an opportunity for cross-validation: having
specifications fail (or pass) both verifiers decreases the amount of
trust placed in a single verifier, as listed in point
\ref{asm:backends-correct} of our correctness assumptions described in
\cref{sec:using-system}.

\newcommand{\specSymb}{S}
\newcommand{\verifSymb}{V}
\newcommand{\totalSymb}{T}

\begin{table}
{ \setlength{\tabcolsep}{3pt}
\begin{tabular}{lrrrrrrrr}
\toprule
            &  \ryzenabbr & \aspeedabbr & \jetsonabbr &   \picoabbr &    \stmabbr &  \txoneabbr &   \omapabbr &   \zynqabbr\\
\midrule
\specSymb   &        1956 &        2504 &        9667 &        1096 &        3984 &        2366 &        1934 &        2319\\
\verifSymb  &         127 &         223 &          61 &         450 &         126 &        1160 &         325 &         193\\
\midrule
\totalSymb  &        2083 &        2727 &        9728 &        1546 &        4110 &        3526 &        2259 &        2512\\
\bottomrule
\end{tabular}
 }
\caption{Lines of code counts for our case studies.\\ \specSymb: specification, \verifSymb: bug-finding and proof scenarios, \totalSymb: total.}
\label{tab:spec-loc-table}
\end{table}

\section{Case studies}
\label{sec:case-studies}

We applied \system to 8 \glspl{soc} for which we had hardware
documentation.
We obtained most documentation by downloading the manuals and additional
resources from the CPU or SoC vendor's websites. These were, in general,
available freely either unauthenticated or after a short registeration
process, with exception of documents for the \txone: we obtained access
to these under a non-disclosure agreement.
We manually created machine-readable
specifications in \system's modelling language and also specified
security properties based on the goals of protection
mechanisms like \tz and the need to protect the integrity of code in
highly privileged environments like x86 \gls{smm}.

On the bug-finding side, we found both documentation
errors and actual hardware bugs, including the previously undiscovered
vulnerability described in \cref{sec:overview}.
For some systems, we also uncovered critical constraints
on secure configuration which were not explicitly mentioned in the
documentation.

On the proof side, we showed for four configurations on three
different platforms that they achieve memory integrity and
confidentiality properties.
In two cases, we were able to model recommendations for secure
configuration from the documentation and prove they achieved the desired
isolation.

\subsection{Bug finding}

We have already introduced the \textbf{\textit{Marvell \txone
(\txoneabbr)}}~\cite{thunderx1-hrm}, a 48-core ARMv8-A \gls{soc} which we used
as an example in \cref{sec:overview}. The full model contains more details, and
includes more devices than the \lstinline{MiniThunderX1}, but exhibits
the same vulnerability due to a missing security check in its \lstinline{ASC}.

\glsreset{smm}

\paragraph{\ryzen (\ryzenabbr)}
The \ryzen series is a recent implementation of x86
\cite{AMD:Ryzen}.  We model several
protection-related aspects of the architecture and the registers that control
it, including \gls{smm}, also called ``ring -2'', a highly privileged
operating mode of x86 cores for functions like power management.

The recently-presented ``Sinkclose''
vulnerability~\cite{sinkclose} is a privilege escalation bug in almost all
recent AMD CPUs where code executing with ring 0 (regular kernel) privileges
can cause the \gls{smm} interrupt handler to jump to instructions outside of
\gls{smm}-protected memory, executing arbitrary code.

Our model captures this vulnerability as a property of the
architectural instruction pointer: at no point during execution of a
\gls{smi} should it point outside of a \gls{smm}-protected region.
Based on our specification of the architecture, \system correctly
finds the vulnerability and outlines an exploit.  Curiously, while the
original authors mention that the vulnerability is documented, we
believe that AMD Ryzen manual incorrectly describes the hardware, and
the vulnerability is only present in older documentation.

\paragraph{\aspeedlong (\aspeedabbr)}
The \aspeed~\cite{ASPEED:ast2600} is a \gls{soc} widely used as a
\gls{bmc} on servers, and supports \tz as isolation
mechanism for separating untrusted, remotely-accessible software from
highly privileged power management firmware.

In developing the \system model for \aspeedabbr, we found
that the built-in \gls{dma}-capable \gls{nic} is always considered a
Secure peripheral by the interconnect, allowing it to read and
write Secure memory.  While not an exploit \emph{per se},
it violates the principle of least-privilege: if the Non-Secure world
were allowed to program the \gls{nic}, it could trivially bypass \tz
protections via the \gls{nic}'s \gls{dma} capabilities.

Using \system, we can test a
configuration that prevents the Non-Secure world from breaking its confinement,
but it requires the \gls{nic} driver to run in the secure world.

\paragraph{\jetsonlong (\jetsonabbr)} The \jetson~\cite{nvidia:jetson} is a
general-purpose \gls{som} for industrial applications combining different ARM
cores with an Nvidia Pascal GPU.

This was the most difficult system to model in \system.  The manual is
highly ambiguous throughout, particularly with regard to how the large
number of hardware components on the \jetson interact.  For example,
firmware can partition DRAM into separate regions in several different
ways: \tz secure memory, a ``Video Protection Region'', and a range of 4KB-aligned
``generalized carveout regions''.  Where these regions overlap, the
access rights are implementation-defined.  \system does determine that
incorrect software configurations might (and in some cases do) bypass \tz.

\paragraph{\picolong (\picoabbr)} The \pico~\cite{rpi:pico2} is a
microcontroller with features found on a larger \gls{soc},
including dual application cores with \tz and a \gls{dma} engine.

Based on the manual, \system detects that the \gls{dma}
engine's security level is set by a privileged but Non-Secure core, which
can bypass \tz isolation (as with \aspeedabbr).  We tested
the inputs that \system's exploit found on a real
\pico and found that Non-Secure cores' writes to the privilege level
were silently ignored, so this is a documentation bug.

\paragraph{\stmlong (\stmabbr)} The \stm~\cite{stm32h7}
provides a hardware feature called proprietary code
readout protection which can make a given
region in flash memory unreadable for regular memory reads and
only accessible via instruction fetches.
The \stm features three
read-out protection levels, and when decreasing the read-out
protection level, the protected region in the flash memory is
mass-erased.
An erratum for the \stm points out that when decreasing the level from 1 to 0,
the protected area may become unprotected.  This behavior is
consistent with the documentation, and indeed based on our model of
the \stm from the original manual \system rediscovers this
bug.

\paragraph{\omaplong (\omapabbr)}
The \omap~\cite{TI:OMAP:2014} is
a multimedia SoC
that was used in smartphones and tablets until recently.  One of its
several interconnects (``L4'') has a series of memory-mapped Address
Protection (AP) registers that control access to the L4 interconnect
itself, and are used to implement firewalls between different devices
and cores.  Until they are programmed correctly, they can be modified
by any actor that can access their addresses, including DMA-capable
devices.
\system does not find any apparent hardware bugs based on modelling
the \omap manual, but does determine that if the AP registers are not
configured properly, the System DMA controller can modify the AP
registers and gain access to address regions that it should not have
access to.

\paragraph{\zynqlong (\zynqabbr)} The
\zynq~\cite{xilinx:ultrascaleplus} is a modern multi-processor \gls{soc}
with an \gls{fpga}.
It supports \tz, using several instances of the region-based
\gls{xmpu}.
The \zynq includes 8 DMA controllers, each of which can be placed in the Secure
or Non-Secure world based on a system-level configuration register
called \lstinline{slcr_gdma}.  This register can only by modified from the Secure
world.  Additional protection against bugs in the software running in
the Secure world is provided by a lock register.
Once this lock register is set, \lstinline{slcr_gdma} cannot be
modified further until reboot.
We model a scenario where the lock is not set, so
a secure DMA channel could be controlled from the Non-Secure
world if software in the Secure world accidentally changes
\lstinline{slcr_gdma}.

\subsection{Proofs}

Modeling system platforms and finding bugs in them is useful, but
we want to go further and \emph{prove} facts about them.

\paragraph{\txoneabbr} We prove that it is possible to correctly use the
\txone's compression acceleration (ZIP) engine, without allowing it to modify
unrelated data in the system. Our model includes the ZIP engine's
descriptor formats, but limits the engine itself to a single channel. The
\gls{smmu} is abstracted: instead of bit-precisely modeling the translation
table format and walk, the \gls{smmu} is reduced to a virtual-to-physical
mapping, and a region of DRAM. To over-approx\-i\-mate the effects of
accesses to the \gls{smmu} translation tables, any write to the translation table
region can arbitrarily change the \gls{smmu}'s mappings. Our top-level statement
is of the form described in English in~\cref{sec:compare-initial-final-snapshots}
and by code in \ref{app:comparing_snapshots}.

\paragraph{\picoabbr} We have proven two statements about configuring the
software on the \pico.
The first one shows that
if the \gls{dma} engine's security level can only be set by the Secure world
(which contradicts the documentation, but seems to be true based on our testing),
then a particular configuration of the \pico platform exists such that
a designated region of secure memory cannot
be accessed by a combination of the \gls{dma} engine and non-secure software
running on core 0 of the \pico.
This proof assumes that core 1 is disabled, which is a common use case:
For example, the Zephyr OS \cite{zephyr-docs} only supports single-core
operation on the \pico.

In the second statement, we also include core 1, which turns out to be
tricky: both cores on the \pico have their own copies of \tz-related \gls{mpu}
registers, and the RP2350 reference manual mentions that instead of
carefully synchronizing the two copies, it may be simpler to use the cores
asymmetrically, using the \lstinline{FORCE_CORE_NS} register to mark all
core 1 accesses Non-Secure on the system bus.
We prove that there exists a configuration of the \pico, using the
\lstinline{FORCE_CORE_NS} register, which protects a region of memory such that
it can only be accessed by Secure software running on core 0. Any other
requester (\gls{dma}, core 1, or Non-Secure code on core 0) is unable to
successfully request an address within the protected region. Furthermore, we
show that there is a sequence of register writes, starting from the initial
post-boot startup state which configures the system in the given manner, thus
also proving that this state is reachable from a fresh boot.

Both proofs use a monitor-based top-level statement as described
in~\cref{sec:read-write-monitor} (see \ref{app:monitor} for a code
example) and rely on an invariant which protects
the integrity of the access control registers.

\paragraph{\zynqabbr} During modeling of the \zynq, we encountered
another recommendation in the technical manual, also concerning isolation of
memory regions using \tz. In order to prevent buggy systems code, or malicious
actors, from circumventing this isolation, the \tz configuration on the \zynq
can be locked using a write-once register. This prevents any modification of
world assignment, even by other secure requesters. Our top-level statement proves that there exists a
configuration of the \zynq platform, such that a region of DRAM can be
isolated indefinitely from other requesters, and these guarantees continue
to hold even in the presence of buggy \tz \gls{dma} drivers,
using a monitor similar to the one in \ref{app:monitor}
as well as a proof by induction.

\subsection{Discussion}

In the course of specifying \glspl{soc}, we found numerous
omissions and ambiguities in reference manuals in addition to the
more serious cases reported above.  This is perhaps unsurprising given
the use of prose to describe something highly precise.
Moreover, the effort of creating a specification is not trivial:
significant work is needed to understand a platform in enough fidelity to build
a suitable model for it.

However, once a
component is modeled, its semantics are available indefinitely. It reveals the
complexity of interactions at key points in modern \gls{soc}, such as
the interconnect: many different resource protection mechanisms
converge at a single point, and it is not immediately clear how they
fit together. A formal model makes the complexity of the underlying system
visible, and forces the specifier to engage with it.

Weighing this against the wasted software engineering effort of dealing with
hardware bugs, documentation errors, and misinterpretations of
technical manuals is beyond the scope of this work, but we expect the
advantages of a formal model with automatic analysis to outweigh the
creation cost.

While our specifications capture arbitrary interleavings of requests
from different initiators, each request (and any others it initiates)
is assumed to run to completion before the next one starts.  While
there are no language-level restrictions in \system to modelling
concurrent behavior, it is cumbersome, and we are exploring extensions
to our DSL to express true concurrency more concisely.

\section{Evaluation}
\label{sec:evaluation}

We use the described scenarios to answer the following questions about \system:
\begin{itemize}
  \item[\cref{sec:expressivity}] Can \system's \gls{dsl} express informal technical reference manuals
    as a formal, machine-reachable model?
  \item[\cref{sec:assertion_expressivity}] Can security properties and proofs be expressed as well?
  \item[\cref{sec:execution-time}] How fast does \system verify proofs and find bugs?
  \item[\cref{sec:spec-time}] How quickly can new users specify a \gls{soc} in \system?
\end{itemize}
We run our performance evaluations on an Arm-based Apple MacBook Pro with a
10-core M1 Pro chip, and 32GB of RAM. We evaluated commit \sockeyeCommit{} of
\system, and the following versions of our backend tools: Z3 4.13.3, CBMC 6.8.0,
and Rosette commit \texttt{29808a02}.

\begin{table}
  \centering
  \begin{threeparttable}
  \begin{tabular}{lrrrrrr}
    \toprule
    &\multicolumn{2}{c}{bug scenario} & \multicolumn{2}{c}{bugfix scenario}\\
    \cmidrule(lr){2-3} \cmidrule(lr){4-5}
    SoC \hspace{16mm} & Rosette & CBMC & Rosette & CBMC\\
    \midrule
    \ryzenabbr & 30.0 & to & 9.6 & to \\
    \aspeedabbr & 542.4 & to & 469.2 & to \\
    \jetsonabbr & 1.1 & 0.4 & 1.1 & 1.1 \\
    \picoabbr & 6.4 & 738.1 & 4.8 & 701.0 \\
    \stmabbr & 7.5 & na$^1$ & 3.4 & na$^1$ \\
    \txoneabbr & 1.2 & 64.4 & 1.1 & 2.6 \\
    \omapabbr & 74.3 & 19.7 & 81.8 & 20.4 \\
    \zynqabbr & 16.1 & na$^2$ & 7.9 & na$^2$ \\
    \bottomrule
  \end{tabular}
  \begin{tablenotes}
    \footnotesize{
    \item[1]causes \texttt{cbmc} to run out of memory.
    \item[2]uses \texttt{BitInt(n)} where $n > 64$ which our C backend does not support.
    }
  \end{tablenotes}
  \caption{Runtime of our bug-finding/bug-fixing scenarios in seconds.
    ``na'' = not applicable, and ``to'' = time-out (>900s)}
  \label{tab:bug-runtimes}
  \end{threeparttable}
\end{table}

\subsection{\system can express the complex interactions on SoCs}
\label{sec:expressivity}

Over the course of modelling the 8 platforms described in
\cref{sec:case-studies}, we found \system's semantics are expressive enough
to capture a diverse set of platforms. We have successfully modeled application
and management cores, \gls{dma} engines and other \gls{dma}-capable devices,
SMMUs/IOMMUs, and other \gls{soc}-specific protection devices.

\system's design forces us to make implicit, potentially ambiguous parts explicit.
For example, tracking the bitwidth of all integer expressions in the typechecker
forces us to be explicit about how values are (sign-)extended or truncated.

On the other hand, if we intentionally want to leave something underspecified,
we can say so using \lstinline{any<T>}.
This allows us to conservatively approximate the set of possible platform
behaviors, with the ability to refine later if necessary.

\subsection{\system's proof language is just expressive enough}
\label{sec:assertion_expressivity}

Compared to interactive proof assistants, \system's assertion language
is much less expressive.
For instance, it is impossible to write properties like
\lstinline{always_holds(desired_property)}
and have such a claim checked by \system in one go.
Instead, as illustrated in \cref{code:example_induction_proof},
we need to run three separate checks,
a \lstinline{base_case} that establishes an \lstinline{invariant},
an \lstinline{inductive_step} that shows that the \lstinline{invariant} is
preserved,
as well as a check that the \lstinline{invariant} implies the
\lstinline{desired_property}.
The final reasoning step needed to conclude that these three checks imply that
the \lstinline{desired_property} always holds needs to be done by the user.
While reviewing each other's \gls{soc} models, it turned out that performing
this final reasoning step is only a small fraction of the general code reviewing
effort, so, pragmatically, it is not a problem that this step is not automatic.

As described in \cref{sec:security_properties}, \system supports various
patterns for proving standard security properties, such as
confidentiality, integrity, and non-interference, most which we have
also successfully applied in practice, despite the lack of $\forall$ and
$\exists$ quantifiers, or the lack of equality testing for vectors.
However, \system in its current form cannot express liveness properties
(properties of the form ``something good eventually happens'').

While we have shown that the \system \gls{dsl} is expressive enough to specify
real-world devices in sufficient fidelity, it is not always
possible to do so succinctly.
Some of \system's users have dealt with this problem
by automatically generating parts of their specification, for example from
structured specifications of register files that some vendors publish.

\begin{table}
  \centering
  \begin{threeparttable}
  \begin{tabular}{lrrrrrr}
    \toprule
    &\multicolumn{2}{c}{base case} & \multicolumn{2}{c}{step case}\\
    \cmidrule(lr){2-3} \cmidrule(lr){4-5}
    Proof & Rosette & CBMC & Rosette & CBMC\\
    \midrule
    \picoabbr (DMA) & 2.3 & 2.6 & 2.8 & to \\
    \picoabbr (TrustZone) & 2.1 & 2.4 & 3.2 & 353.3 \\
    \txoneabbr (ZIP) & 0.7 & 59.0 & 0.7 & 354.6 \\
    \zynqabbr (TrustZone) & 1.1 & na$^1$ & 15.1 & na$^1$ \\
    \bottomrule
  \end{tabular}
  \begin{tablenotes}
    \footnotesize{
    \item[1]uses \texttt{BitInt(n)} where $n > 64$ which our C backend does not support.
    }
  \end{tablenotes}
  \caption{Runtime of our proofs in seconds.
    ``na'' = not applicable, and ``to'' = time-out (>900s)}
  \label{tab:proof-runtimes}
  \end{threeparttable}
\end{table}

\subsection{\system finds hardware and documentation bugs and correctness proofs
in reasonable time}
\label{sec:execution-time}

\Cref{tab:bug-runtimes,tab:proof-runtimes} show solving performance
for the bugs and proofs of the \gls{soc} specifications described in
\cref{sec:case-studies}.
We observe that most scenarios complete fast enough for interactive development.
In general, the Rosette backend
outperforms CBMC: we suspect the C code generated by \system might not be
particularly amenable to CBMC, and we have not explored the space of possible
CBMC configuration flags yet.
It is well-known that fully-automatic verification tools are performance
sensitive even to seemingly small or unrelated changes. Thus, having the option
to run a query through multiple different tools by some sort of
portfolio-solving approach is useful.

When facing a time-out, users engaging in bug-finding have the choice of
encoding domain-specific assumptions or restricting the attack search
space to speed up solving times, for example by offering fewer
non-deterministic choices.

For instance, when trying to run the most general bug scenario described
for the \zynqabbr, Rosette times out. However, if we restrict the DMA
channel to a specific one, the bug is found within 2~minutes, and if the
solver helped further by fixing one step to be a step where the CPU
initiates a DMA transfer, the bug is found in less than 20~seconds.

A similar approach can be taken for the proofs as well: providing assertions can
change the order of heuristics of solver backends apply, and
dramatically improve solving times. We have employed this approach for
proving how to correctly configure the compression accelerator on the
Marvell \txone.
It is worth noting that proofs generally run
faster than bug-finding scenarios, because inductive proofs only need to
reason about a single system step at a time.

\subsection{New users can pick up \system within months}
\label{sec:spec-time}

Many of our existing specifications were written by bachelor's and
master's-level students as part of individual research projects or theses.
Most of the students had little prior experience with formal modelling and
\glspl{soc}. With a bit of guidance, they were able to start modelling their
individual project within a few weeks of their theses, and found that the main
limitation is not the \system language, but rather the complexity and
underspecification of their platform's reference manuals.

All students managed to model bugs and bugfixes. However, writing meaningful
proofs can be more involved, as witnessed by the fact that
\cref{tab:proof-runtimes} has fewer rows than \cref{tab:bug-runtimes}.

\section{Related Work}

\paragraph{\gls{soc} address modelling}
Decoding nets were introduced by Achermann \etal~\cite{achermann:itp:2018}.
Their static representation captures the system at a single point in time.
In contrast, our approach supports dynamic updates to the address translation
state. %

\paragraph{MMU configuration and driver verification} Velosiraptor~\cite{achermann:asplos:2025}
formalizes the semantics of \gls{mmu} and other memory protection hardware, and
synthesizes operating system code to program them.
This answers the question of what bit patterns to write to protect a memory region,
but leaves the question which regions need to be protected unanswered,
which our work focuses on. Termite~\cite{Termite_SOSP09} and
Pancake~\cite{zhao:arxiv:2025} introduce driver synthesis for individual
devices, but their models lack the capability to represent how
these devices affect the interactions on \glspl{soc} as a whole.

\paragraph{\gls{soc} analysis} Existing work on \gls{soc} verification and
fuzzing assumes access to the \gls{soc}'s \gls{rtl}
code~\cite{SoC_Confidentiality_DAC21,hossain:date:2023}. In contrast,
our approach is applicable to any \gls{soc} with a suitable reference manual,
though the quality of the manual plays a key role in how useful the
resulting specification will be.

\paragraph{\gls{tee} analysis}
Previous works on verified \gls{tee} verification, such as
Komodo~\cite{ferraiuolo:sosp:2017}, Cerberus~\cite{lee:ccs:2022},
ProveriT~\cite{hu:ieeedsc:2024}, and Arm CCA~\cite{li:osdi:2022} focus on the
reference monitor and its interaction with a single core, and assume integrity
and/or confidentiality of memory regions used in \gls{tee} operation. Our work
enables formally stating and verifying those assumptions: that other components
cannot interfere with the \gls{tee} monitor.

\paragraph{Hardware verification}
Existing approaches for hardware \emph{implementation}
verification~\cite{witharana:acm:2022} assume access to the underlying \gls{rtl}
description. Not only do systems programmers rarely have access to the
source code of the platform they are programming for, but they care more
about the \emph{software-visible} hardware behavior rather than
\gls{rtl} implementation details.

Reid's work on formalizing and validating Armv8 architecture
profiles~\cite{reidTrustworthySpecificationsARM2016,reidWhoGuardsGuards2017a}
describes the path to a functional specification for architectural behavior of
individual processing units. Interactions with other devices present on
an \gls{soc} cannot be part of the architecture specification, but
the architecture specification could be integrated into \system models in the
future (as described in \cref{sec:future-work}).

\paragraph{Side channels} Our proposed approach focuses on modelling the direct
interactions between components. We consider work on side
channel mitigation~\cite{zhang:acm:2024} an important, but orthogonal
component.
Pensieve~\cite{Pensieve_ISCA23} uses an approach based on Rosette~\cite{Rosette_PLDI14} similar to ours,
but for finding microarchitectural side channels.
However, hand-writing untyped Rosette code like they do,
as opposed to generating Rosette from a typed language like \system,
comes with serious usability drawbacks:
Since Rosette is untyped, trivial errors such as comparing signals of different
bit-widths or passing the wrong number of arguments to a function cause the same
kind of assertion failure as actual bugs and involve considerable debugging effort to
nail down.
In contrast, \system provides located typechecking errors, and by
re-running the program with the values of the counterexample found by
Rosette, it can tell which assertion failed and provide a precise
function-call trace of the exploit.

\section{Future work}
\label{sec:future-work}

As we saw in \cref{sec:evaluation}, the different backends exhibit
the high running time variance that is expected of automatic program verification.
Supporting more backends, such as e.g. KLEE~\cite{cadar:osdi:2008},
and invoking all of them in parallel, might therefore decrease the wall-clock
time elapsed before the user gets a result.

For the bug-finding side, we could also explore \emph{fuzzing}, either
using off-the-shelf tools such as AFL++~\cite{fioraldi:woot:2020} or a custom
fuzzer for \system using LibAFL~\cite{fioraldi:ccs:2022}.

By post-processing the exploits found by \system, it should be feasible to
automatically generate code that sets up a real piece of hardware and tests
the exploit.
Or, turning it the other way, it could be useful to ingest an existing platform
\emph{configuration} by reading out critical registers and memory locations, and
analyzing its security guarantees.

Currently, the only way to produce a new specification of some piece of hardware
is to transcribe its manual by hand, which takes signficant time and effort.
Automatic generation of \system specifications could
significantly speed up this procedure.
Where formal specifications in another language are already available (e.g.
the Arm Architecture Specification
Language~\cite{reidTrustworthySpecificationsARM2016}, or the Sail architecture
specification language~\cite{armstrong:acm:2019}),
we could implement translators from these languages to \system.
Where no formal specifications are available yet, we could also explore using LLMs
to translate the English specifications into \system.

\section{Conclusion}
\label{sec:conclusion}

We have demonstrated that \system can be used to formalize hardware reference
manuals for a set of eight diverse hardware platforms. Using \system's
analysis backends, we found documentation and hardware bugs and
proved security properties for four configurations of three platforms. During
this process, we discovered a previously unknown vulnerability on the
\txone and the related OCTEON family of CPUs.

Using \system for writing platform specifications forces us
to consider exactly what goes on in a modern platform, and requires us to
spell out implicit information often omitted.
Once we have decided on
an interpretation, the analysis backends enable us to automatically
analyze and reason about the (in-)security of our specifications.

\bibliographystyle{ACM-Reference-Format}
\bibliography{references}

\appendix

\clearpage

\section{Full source code of minimized example}\label{app:mini_thunderx1}

Note that the material in this appendix has not been peer-reviewed. 

\begin{lstlisting}
// control if we want the buggy or non-buggy version
const BUGGY: Bool = any;

type PhysAddr = BitInt(48);
type Request = {
  is_write: Bool,
  is_secure: Bool,
  address: PhysAddr,
  value: BitInt(64)
};

type Response = {
  ok: Bool,
  value: BitInt(64)
};

module Region {
  instance START: State<BitInt(64)>(0);
  instance END: State<BitInt(64)>(0);
  instance ATTR: State<BitInt(64)>(0);

  fn allows(request: Request) -> Bool {
    START.get()[47 downto 0] <= request.address &&
    request.address <= END.get()[47 downto 0] &&
    ATTR.get()[if request.is_secure { 1 } else { 0 }]
  }

  fn get(i: BitInt(2)) -> Response {
    match i {
      0 => { ok: true, value: START.get() },
      1 => { ok: true, value: END.get() },
      2 => { ok: true, value: ATTR.get() },
      3 => { ok: false, value: any }
    }
  }

  mut fn set(i: BitInt(2), val: BitInt(64)) -> Response {
    let is_ok_addr = 
      val[2 downto 0] == 0 && val[63 downto 48] == 0;
    let is_ok_attr = val <= 2;
    match i {
      0 => if is_ok_addr {
        START.set(val); { ok: true, value: any }
      } else { { ok: false, value: any } },
      1 => if is_ok_addr {
        END.set(val); { ok: true, value: any }
      } else { { ok: false, value: any } },
      2 => if is_ok_attr {
        ATTR.set(val); { ok: true, value: any }
      } else { { ok: false, value: any } },
      3 => { ok: false, value: any }
    }
  }

  mut fn havoc() {
    START.set(any);
    END.set(any);
    ATTR.set(any);
  }
}

module DRAM {
  // 2^31 8-byte words = 2^34 bytes = 16GB
  instance storage: Array<BitInt(31), BitInt(64)>;

  fn truncate_addr(addr: PhysAddr) -> BitInt(31) {
    addr[33 downto 3]
  }

  fn load(a: PhysAddr) -> Response {
    let v = storage.load(truncate_addr(a));
    printf("DRAM: Loading {v} from {a}\n");
    { ok: true, value: v }
  }

  mut fn store(a: PhysAddr, v: BitInt(64)) -> Response {
    printf("DRAM: Storing {v} to {a}\n");
    storage.store(truncate_addr(a), v);
    { ok: true, value: any<BitInt(64)> }
  }

  mut fn havoc() {
    storage.set(any);
  }
}

module ASC {
  instance region0: Region;
  instance region1: Region;
  instance region2: Region;
  instance region3: Region;

  mut fn config_write(
    region_id: BitInt(2),
    register_id: BitInt(2),
    value: BitInt(64)
  ) -> Response {
    printf("ASC: Setting region{region_id}.");
    match register_id {
        0 => printf("START to {value}\n"),
        1 => printf("END to {value}\n"),
        2 => printf("ATTR to {value}\n"),
        3 => printf("??? to {value}\n"),
    };
    match region_id {
      0 => region0.set(register_id, value),
      1 => region1.set(register_id, value),
      2 => region2.set(register_id, value),
      3 => region3.set(register_id, value),
    }
  }

  fn config_read(
    region_id: BitInt(2),
    register_id: BitInt(2)
  ) -> Response {
    match region_id {
      0 => region0.get(register_id),
      1 => region1.get(register_id),
      2 => region2.get(register_id),
      3 => region3.get(register_id),
    }
  }

  callee dram: DRAM;

  fn is_region_config_addr(a: PhysAddr) -> Bool {
    a[47] && a[46 downto 7]==0 && a[2 downto 0]==0
  }

  fn is_allowed_dram_addr(r: Request) -> Bool {
    r.address < 1 << 34 &&
    r.address[2 downto 0] == 0 &&
    (region0.allows(r) || region1.allows(r) ||
     region2.allows(r) || region3.allows(r))
  }

  mut fn asc_region_request(r: Request) -> Response {
    let region_id = r.address[6 downto 5];
    let register_id = r.address[4 downto 3];
    if r.is_write {
      config_write(region_id, register_id, r.value)
    } else {
      config_read(region_id, register_id)
    }
  }

  mut fn dram_request(r: Request) -> Response {
    if r.is_write {
      dram.store(r.address, r.value)
    } else {
      dram.load(r.address)
    }
  }

  mut fn request(r: Request) -> Response {
    if is_region_config_addr(r.address)
      && if BUGGY { true } else { r.is_secure }
    {
      asc_region_request(r)
    } else if is_allowed_dram_addr(r) {
      dram_request(r)
    } else {
      { ok: false, value: any<BitInt(64)> }
    }
  }

  mut fn havoc() {
    region0.havoc();
    region1.havoc();
    region2.havoc();
    region3.havoc();
  }
}

module CPU {
  callee asc: ASC;
  instance is_secure: State<Bool>(true);

  mut fn step() {
    let r: Request = {
      is_write: any<Bool>,
      is_secure: is_secure.get(),
      address: any<PhysAddr>,
      value: any<BitInt(64)>
    };
    printf("CPU: request is {r}\n");
    let ignored_reply = asc.request(r); ()
  }

  mut fn store_phys(a: PhysAddr, v: BitInt(64)) {
    let ignored_reply = asc.request({
      is_write: true,
      is_secure: is_secure.get(),
      address: a,
      value: v
    }); ()
  }

  mut fn havoc() {
    is_secure.set(any);
  }
}

module MiniThunderX1 {
  instance cpu: CPU;
  instance asc: ASC;
  instance dram: DRAM;
  asc.dram -> dram;
  cpu.asc -> asc;

  mut fn step() {
    // in a more complete model, more components could
    // take a step, e.g. a DMA engine, NIC, ...
    cpu.step()
  }

  mut fn havoc() {
    cpu.havoc();
    asc.havoc();
    dram.havoc();
  }
}

module Main {
  instance miniTX1: MiniThunderX1;

  const CONFIG_BASE: PhysAddr = 1u48 << 47;
  const ATTR_EMPTY: BitInt(64) = 0;
  const ATTR_NONSEC: BitInt(64) = 1;
  const ATTR_SEC: BitInt(64) = 2;

  mut fn setup_region(
    region_id: BitInt(2),
    start: BitInt(64),
    end: BitInt(64),
    attr: BitInt(64)
  ) {
    let a = CONFIG_BASE + (0u41 ## region_id ## 0u5);
    miniTX1.cpu.store_phys(a, start);
    miniTX1.cpu.store_phys(a + 8, end);
    miniTX1.cpu.store_phys(a + 16, attr);
  }

  mut fn setup_regions() {
    // First 4MB are secure
    setup_region(0, 0, (1<<24)-1, ATTR_SEC);
    // Everything else up to 16GB is non-secure
    setup_region(1, 1<<24, (1<<34)-1, ATTR_NONSEC);
    // Regions 2 and 3 are unused
    setup_region(2, 0, 0, ATTR_EMPTY);
    setup_region(3, 0, 0, ATTR_EMPTY);
  }

  mut fn test_secure_area_unchanged() {
    setup_regions();
    miniTX1.cpu.is_secure.set(false);
    let orig_mem = miniTX1.dram.storage.get();
    miniTX1.step();
    miniTX1.step();
    let new_mem = miniTX1.dram.storage.get();
    let test_addr = any<BitInt(31)>;
    assume(test_addr <= 0x1f_ffffu31);
    assert(orig_mem[test_addr] == new_mem[test_addr])
  }

  fn disjoint_ranges(
    start1: PhysAddr, end1: PhysAddr,
    start2: PhysAddr, end2: PhysAddr
  ) -> Bool {
    end1 < start2 || end2 < start1
  }

  fn region_doesnt_allow_nonsecure(
    region_id: BitInt(2),
    range_start: PhysAddr,
    range_end: PhysAddr
  ) -> Bool {
    let start = miniTX1.asc.config_read(region_id, 0);
    let end   = miniTX1.asc.config_read(region_id, 1);
    let attr  = miniTX1.asc.config_read(region_id, 2);
    attr.value == ATTR_EMPTY ||
    attr.value == ATTR_SEC ||
    disjoint_ranges(
      start.value[47 downto 0], end.value[47 downto 0],
      range_start, range_end)
  }

  fn invariant() -> Bool {
    region_doesnt_allow_nonsecure(0, 0, (1<<24)-1) &&
    region_doesnt_allow_nonsecure(1, 0, (1<<24)-1) &&
    region_doesnt_allow_nonsecure(2, 0, (1<<24)-1) &&
    region_doesnt_allow_nonsecure(3, 0, (1<<24)-1) &&
    !miniTX1.cpu.is_secure.get()
  }

  mut fn base_case() {
    setup_regions();
    miniTX1.cpu.is_secure.set(false);
    assert(invariant());
  }
  mut fn inductive_step() {
    assume(!BUGGY);
    miniTX1.havoc();
    assume(invariant());
    miniTX1.step();
    assert(invariant());
  }

  mut fn invariant_is_useful() {
    miniTX1.havoc();
    assume(invariant());
    let orig_mem = miniTX1.dram.storage.get();
    miniTX1.step();
    let new_mem = miniTX1.dram.storage.get();
    let test_addr = any<BitInt(31)>;
    assume(test_addr <= 0x1f_ffffu31);
    assert(orig_mem[test_addr] == new_mem[test_addr])
  }
}
\end{lstlisting}

\clearpage

\section{Source snippets illustrating different kinds of properties}\label{app:sample_properties}

Note that the material in this appendix has not been peer-reviewed. 

\newcommand{\propertyKindTitle}[1]{\subsection{#1}}

\propertyKindTitle{Directly testing the output of an access-control function}\label{app:directly_testing}

\begin{lstlisting}
module Main {
  instance acu: AccessControlUnit;

  fn is_acu_correct() -> Bool {
    acu.can_access(SecureCPU, 0x0, 0xFF) &&
    !acu.can_access(NonSecureCPU, 0x0, 0xFF)
  }

  fn test_acu_after_two_steps() {
    acu.do_initial_setup();
    assert(acu_is_correct());
    step(); step();
    assert(acu_is_correct());
  }
}
\end{lstlisting}

\propertyKindTitle{Comparing initial and final state snapshots}\label{app:comparing_snapshots}

\begin{lstlisting}
  fn test_secure_ram_cannot_be_modified() {
    setup();
    let orig_sec_ram = ram.get()[0x3000..0x3FFF];
    step(); step();
    let new_sec_ram = ram.get()[0x3000..0x3FFF];
    let test_addr = any<BitInt(12)>;
    assert(orig_sec_ram[test_addr]
             == new_sec_ram[test_addr]);
  }
\end{lstlisting}

\propertyKindTitle{Specifying that an area cannot be read using a monitor}\label{app:monitor}

\begin{lstlisting}
module DramMonitor {
  instance got_bad_request: State<Bool>(false);
  callee dram: Dram;
  fn handle_request(req: Request) -> Response {
    got_bad_request.set(
      got_bad_request.get() ||
      (is_secure_address(req.addr) &&
       !is_secure_originator(req.orig))
    );
    dram.handle_request(req)
  }
}
module Main {
  instance dram: Dram;
  instance dram_monitor: DramMonitor;
  dram_monitor.dram -> dram;

  fn test_secure_ram_cannot_be_modified() {
    setup(); step(); step();
    assert(!dram_monitor.got_bad_request());
  }
}
\end{lstlisting}

\propertyKindTitle{Expressing integrity as preservation of an invariant}\label{app:integrity_by_invariant}

\begin{lstlisting}
module Main {
  instance acu: AccessControlUnit;
  fn nontrivial_invariant() -> Bool {
    acu.get_setting1() == value1 &&
    ...
    acu.get_settingN() == valueN
  }
  fn base_case() {
    havoc();
    setup();
    assert(nontrivial_invariant());
  }
  fn step_case() {
    havoc();
    assume(nontrivial_invariant());
    step();
    assert(nontrivial_invariant());
  }
}
\end{lstlisting}

\propertyKindTitle{Expressing noninterference/information-flow properties using two copies of the \gls{soc}}\label{app:two_copies}

\begin{lstlisting}
module System {
  instance ram: Array<BitInt(12), QuadWord>
  fn assume_low_equivalent(other: System) {
    // ram addresses below 0x300 are high-confidential
    ram.set(any ++ other.ram.get()[0x300..0xFFF]);
    low_register0.set(other.low_register0.get());
    ...
  }
  fn assert_low_equivalent(other: System) {
    let test_addr = any<BitInt(12);
    assume(test_addr >= 0x300);
    assert(ram.load(test_addr) == other.ram.load(test_addr))
    ...
  }
}
module Main {
  instance sys0, sys1: System;
  fn base_case() {
    sys0.assume_low_equivalent(sys1);
    setup();
    sys0.assert_low_equivalent(sys1);
  }
  fn step_case() {
    sys0.assume_low_equivalent(sys1);
    step();
    sys0.assert_low_equivalent(sys1);
  }
}
\end{lstlisting}

\propertyKindTitle{Expressing integrity and non-interference as equivalence with an air-gapped system}\label{app:air_gapped}

\begin{lstlisting}
module ProofBench {
  // The more realistic model
  instance system: System;

  // The spec model, more obviously isolated
  instance p0: ProcessSpec;
  instance p1: ProcessSpec;

  mut fn step(is_context_switch: Bool) {
    // Stepping the intertwined state model
    system.step(is_context_switch);

    // Stepping the spec model
    // Reads system.current_pid to know the schedule.
    if is_context_switch {
      // nothing to be done in the individual procs
    } else {
      if system.current_pid.get() ==  0 {
        p0.step();
      } else {
        p1.step();
      }
    }
  }

  mut fn assert_reads_correctly() {
    let test_vaddr: VirtAddr = any;
    assume(is_valid_vaddr(test_vaddr));
    let expected = if system.current_pid.get() == 0 {
      p0.virtual_mem.load(test_vaddr)
    } else {
      p1.virtual_mem.load(test_vaddr)
    };
    let actual = system.ram.load(
        system.translate(false, test_vaddr).paddr);
    log_tested_values(test_vaddr, expected, actual);
    assert(expected == actual);
  }

  mut fn step_and_check() {
    let is_context_switch: Bool = any;
    step(is_context_switch);
    assert_reads_correctly();
  }
}

module Main {
  instance bench: ProofBench;

  mut fn test_one_step() {
    bench.init_with_zero();
    bench.step_and_check();
  }
}
\end{lstlisting}

\end{document}